\documentclass[11pt,twoside]{book} 
\usepackage{asp2008s}
\usepackage{times}
\usepackage{lscape}
\usepackage{epsf}

\usepackage{graphicx}
\usepackage{amssymb}
\usepackage{longtable}
\usepackage[figuresright]{rotating}
\usepackage{multirow}

\newcommand{\simless}{\mathbin{\lower 3pt\hbox {$\rlap{\raise 5pt\hbox{$\char'074$}}\mathchar"7218$}}}

\mainmatter

\newlength{\deftabcolsep}
\begin{document}

\title{Star Formation and Molecular Clouds \\at High Galactic Latitude}
\author{Peregrine M. McGehee}
\affil{Infrared Processing and Analysis Center, \\
California Institute of Technology, \\
Pasadena, CA 91125 USA}

\begin{abstract}
In this chapter we review the young stars and molecular clouds found
at high Galactic latitudes $(|b| \ge 30^\circ)$. These are mostly
associated with two large-scale structures on the sky, the Gould Belt
and the Taurus star formation region, and a handful of molecular
clouds including MBM 12 and MBM 20 which, as a population, consist of
the nearest star formation sites to our Sun. There are also a few
young stars that are found in apparent isolation far from any
molecular cloud.  The high latitude clouds are primarily translucent
molecular clouds and diffuse Galactic cirrus with the majority of them
seen at high latitude simply due to their proximity to the Sun.  The
rare exceptions are those, like the Draco and other intermediate or
high velocity clouds, found significantly above or below
the Galactic plane.  We review the processes that result in star
formation within these low density and extraplanar environments as
well as the mechanisms for production of isolated T Tauri stars. We present
and discuss the known high-latitude stellar nurseries and young
stellar objects.
\end{abstract}

\section{Introduction}

The young stellar populations at high Galactic latitude
$(|b| \ge 30^\circ)$ are predominately associated with two
structures, the Taurus-Auriga
star formation region and the Gould Belt, that each subtend large
areas on the sky.
Pre-main-sequence stars have been verified in a handful of, generally the
densest, high-latitude molecular clouds: MBM 12 \citep{luh01},
MBM 20 \citep[L1642;][]{san87}, and possibly MBM 33
and MBM 37 \citep{mar96}. A few other high-latitude molecular clouds such
as the MBM 18/MBM 19 complex \citep{cha04}
are suspected of harboring active star formation.

\citet{mag00} deduce on statistical grounds a Gaussian
scale height for the high latitude clouds [HLC] of $\sim$100 pc, consistent
with that determined for lower latitude dark clouds. As \citet{mag96}
note, the HLCs have characteristics that set them apart from the
dark clouds, namely lower densities and isotopic CO abundances
but higher W($^{12}$CO)/W($^{13}$CO) ratios.

The nature of HLCs, except for those having systemic velocities
not consistent with differential Galactic rotation, can be understood in the
context
of the evolutionary scenario outlined by
\citet{sta05}. The origins of the intermediate [IVC]
and high velocity clouds [HVC], as these exceptions are known, are a
matter of significant study.

The initial stage in the formation of molecular clouds
is that atomic gas clouds having a velocity dispersion
$\sim$12 km s$^{-1}$
and scale height $\sim$100 pc condense from the
diffuse atomic gas. As the condensation process continues, the cloud cores
become increasingly molecular resulting in low mass ($< 100 M_\odot$)
high latitude clouds characterized by a scale height of $\sim$70 pc.
The next stage in the cloud growth is the transition,
at masses between 100 and 10$^5$ $M_\odot$, to fully
molecular clouds that have similar
scale heights as the partially molecular high latitude clouds and are capable
of star formation. \citet{sta05} found that in their analysis of the
Bell Laboratories ${}^{13}$CO Milky Way survey the scale height
for small molecular clouds ($M < 2 \times 10^5 M_\odot$) is roughly
independent
of cloud size. For the giant molecular clouds [GMC], however,
the scale height decreases with mass, with the largest clouds always found
in the Galactic mid-plane. This reduced scale height for the GMCs is
attributed to the dissipative
process of GMC formation associated with passage of a spiral arm.

In this formation scenario all but the very nearest molecular clouds
are viewed in projection
against the Galactic plane. The high-latitude clouds are therefore either
very close to the Sun or high above the Galactic plane, in both cases
they warrant careful study of their ability to undergo star formation.
Based on Jeans' mass arguments it is clear that these low density
clouds provide for difficult star formation environments via gravitational
collapse. It should be noted, however, that dense molecular cores have
been detected even in Galactic cirrus, e.g. MCLD 123.5+24.9 \citep{hei02}.
These nearby clouds are ideal laboratories
for study of triggered star formation, for example from supernovae or
stellar winds \citep{elm98}.

In this chapter we begin with the classification of and surveys for
high-latitude molecular clouds. In Sect.~3 we discuss the
high-latitude PMS stars and summarize a recent inventory of the SIMBAD
database.  The individual star formation regions are presented in
Sect.~4. Finally, in Sect.~5 we comment on the environment and
processes of high-latitude star formation.

\section{Molecular Cloud Surveys}

\subsection{Classification of Molecular Clouds}

Molecular clouds
are classed, following \citet{van88}, into three categories based primarily
on the visual extinction ($A_V$) along the line of sight through the cloud.
Diffuse clouds are identified by $A_V < 1$ magnitude, dark clouds are
marked by $A_V > 5$ magnitudes, and translucent clouds are those with
intermediate values of $A_V$. This set of criteria also reflects the
changes in astrochemistry \citep{van88}. While the chemistry in diffuse
clouds is largely due to photoprocesses and that in dark clouds is
dominated by collisional processes, in the translucent clouds we
find the conditions in which carbon is being bound up in the form of CO.

The formation of a molecular cloud requires a column density large enough
to shield H$_2$ and CO from the ambient Galactic ultraviolet flux.
\citet{har01} explain that this protection from disassociation is a
necessary but not sufficient condition for the existence of molecular
material. Other factors that influence the rate of formation of molecular
gas include the gas density, the dust temperature, and pressure forces.
The minimum column density required for shielding is
$N_{\rm H} \sim$ 1--2 $\times 10^{21}$ cm$^{-3}$, corresponding to
 $A_V \sim$ 0.5--1.
While the molecular gas content of diffuse clouds is difficult to detect
in emission due to
low CO column densities \citep[$\le 10^{15}$ cm$^{-2}$;][]{mag96},
optical and UV absorption of CH, CH$^+$, CN, CO, and H$_2$ has been
used historically to determine distances and probe physical conditions
of these objects \citep{wil34,boh78,van89,pen00}.
These absorption line studies have been limited
to sightlines of early spectral type (generally O, B, and A) stars due to
their nearly featureless spectra.

\subsection{Surveys for High-Latitude Clouds}

High-latitude clouds are predominately diffuse or translucent and are
thus difficult to detect on photographic surveys.  The notable
exceptions are the Lynds Dark Nebulae identified by \citet{lyn62}
using Palomar Sky Survey plates covering the sky north of $\delta =
-33^{\circ}$, and extended regions of faint ($SB_V > 25$ mag
arcsec$^{-2}$) nebulosity seen in deep imaging surveys.  The latter
were first identified on the Palomar Sky Survey plates by
\citet{lyn65} and in an un-published study (circa 1968) by
C.~R. Lynds.  \citet{san76} subsequently showed that interpretation of
these structures as reflection nebulosities illuminated by the
Galactic plane yielded surface brightness predictions deduced from the
neutral hydrogen column density that were consistent with
observations. Deep $BRI$ imaging by \citet{guh89} suggest a more
complex scenario since although the $B$ band surface brightnesses
supported the scattered starlight from dust model, the $B-R$ and $R-I$
colors were generally 0.5-1.0 magnitudes redder than predicted.  They
interpret this as due to a broad luminescence feature in small
hydrogenated carbon grains.

The most effective searches for high-latitude clouds have
been based on CO and far infra-red (FIR) dust emission.
Early millimeter surveys for molecular clouds using the 115 GHz
CO (1-0) transition \citep{mag85, ket86, hei88} were based on previously
identified extinction regions or \ion{H}{I} clouds.
\citet{mag96} provided a compilation based on the literature
giving properties and references for clouds with the acronyms UT, ir,
HSVMT, G, Stark and 3C. The HRK catalog \citep{hei88} includes
clouds in regions studied in \ion{H}{I}, CO and the IR.
Subsequent blind all-sky CO (1-0) surveys were performed by \citet{har98}
and \citet{mag00}
for the north and south Galactic hemispheres, respectively.

\citet{des88} combined IRAS 100 $\mu$m data
and the Berkeley \ion{H}{I} survey \citep{hei74}
to detect 516 infrared excess clouds (IREC) at $|b| > 5^{\circ}$.
\citet{rea98} subsequently created higher-resolution maps based on
DIRBE/COBE FIR imaging and the Leiden-Dwingeloo HI survey \citep{har97}
finding 81 new clouds (DIR - Diffuse Infrared Clouds).
\citet{dev87}
conducted a search for FIR emission excess indicative of
the presence of H$_2$ in cold clouds within the Ursa Major region.

High latitude clouds identified by FIR excess emission and
having $E(B-V) > 0.15$ magnitudes are commonly detected in CO(1-0)
\citep{cha06} but not at lower reddenings, thus FIR and CO surveys are
sensitive to different cloud populations.
Assuming $R_V$ = 3.1, this value of $E(B-V)$ corresponds to $A_V \sim$ 0.5
mag which is well below the traditional $A_V \sim$ 1 boundary between
diffuse and translucent clouds thought to mark the onset of CO formation.
The ability of high latitude clouds to more readily form
molecular cores is due to the decreased ambient UV interstellar field.
In the high latitude environment the UV field is thought to be
primarily from
extragalactic sources and
a factor of 2 or 3
lower than the standard value in the solar neighborhood of
$(1-2) \times 10^5$ photons cm$^{-2}$ s$^{-1}$ {\AA}$^{-1}$
over 1000-2000 {\AA}. \citet{hei02} have detected CO(1-0) emission
from molecular cores in
regions having $A_V$ as low as ${\sim}$0.2 magnitudes.
As found by \citet{mag03} and \citet{cha06} the correlation between
$W(CO$) and $E(B-V)$ breaks down at low values of $E(B-V)$. Possible
mechanisms for this lack of correlation include the increased contribution of
dust associated with
atomic hydrogen and local variations in grain properties and the
radiation field.

The major complexes found in the northern Galactic hemisphere
include the Ursa Major cloud complex, which may be physically related to the
Polaris Flare as infrared and 21 cm \ion{H}{I} observations
reveal a common surrounding arc of gas and dust \citep{mey91},
the Draco Intermediate Velocity Clouds \citep[MBM 41--44;][]{goe83},
and the MBM 34 -- MBM 37 clouds associated with the L134N dark cloud
north of the $\rho$ Ophiuchus star formation
region. The southern Galactic sky contains two large structures, the first
being the clouds south ($145^{\circ} \le l \le 195^{\circ},
-46^{\circ} \le b \le -30^{\circ}$) of the Taurus-Auriga star formation
region [SFR], the second is the
MBM 53-55 complex. The southern Taurus clouds may be unrelated to the
Taurus-Auriga SFR due to kinematic considerations \citep{mag00}; however
both these clouds and the L134N region are projected near to the Gould
Belt.

\subsection{Cataloged High Latitude Clouds}

In the recent compendium of Galactic dust clouds by \citet{dut02} only
439 out of 5004 (9\%) are found at
Galactic latitudes $|b| \ge 30.0$.
The \citet{dut02} catalog is derived from a heterogeneous set of surveys
based on IR, CO, and optical detection methods. For the high-latitude
clouds the primary sources used, along with their catalog designations
and detection techniques, are
\citet{mag85} [MBM; CO],
\citet{ket86} [KM; CO],
\citet{des88} [IREC; \ion{H}{I} and FIR],
\citet{mag96} [various; literature], and
\citet{rea98} [DIR, \ion{H}{I} and FIR].
Within this high-latitude sample the majority (393, or 90\%) have central
extinctions less than 1 $A_V$ and are therefore classed as diffuse
clouds. In the \citet{dut02} catalog we find that Lynds 1457, associated with
MBM 12, is the sole high-latitude dark cloud, as defined by $A_V > 5$.
The remaining 45 clouds (10\%) are classed as translucent. The complete set
of high-latitude clouds from \citet{lyn62} are listed in Table 1 where the
$A_V$ extinctions are computed from the ``$E(B-V)$cen'' values of
\citet{dut02} assuming $R_V = 3.1$.

\begin{table}
\caption{Lynds Dark Clouds at $|b| \ge 30^{\circ}$}
\smallskip
\begin{center}
{
\small
\begin{tabular}{lllll}
\tableline
\noalign{\smallskip}
Lynds Number & $l$ & $b$ & Molecular & $A_V$ \\
& [deg] & [deg] &  Cloud & \\
\noalign{\smallskip}
\tableline
\noalign{\smallskip}
L134 & 4.18 & +35.75 & MBM 36 & 2.9 \\
L169 & 5.28 & +36.77 & MBM 37 & 1.7 \\
L183, L184 & 5.70 & +36.62 & MBM 37 & 2.1 \\
L1311 & 125.22 & +32.25 & IREC 167 & 1.0 \\
L1312 & 126.60 & +32.18 & IREC 167 & 0.4 \\
L1453 & 158.66 & $-$33.80 & MBM 12 & 3.8 \\
L1454 & 158.90 & $-$33.77 & MBM 12 & 4.1 \\
L1457 & 159.11 & $-$34.45 & MBM 12 & 5.9 \\
L1458 & 159.14 & $-$33.64 & MBM 12 & 4.2 \\
L1569 & 189.54 & $-$36.67 & MBM 18 & 1.5 \\
L1642 & 210.89 & $-$36.51 & MBM 20 & 2.0 \\
L1778 & 358.97 & +36.82 & MBM 33 & 2.0 \\
L1780 & 359.23 & +36.60 & MBM 33 & 1.7 \\
\noalign{\smallskip}
\tableline
\end{tabular}
}
\end{center}
\end{table}

The asymmetry between the numbers of northern and southern clouds
suggests that
these are primarily a local population of objects.
In the catalog of \cite{dut02} we see that 54\% of the
high latitude diffuse clouds and 63\% of the high latitude
translucent and dark clouds are found in the Galactic Southern hemisphere.
By assuming that
the molecular clouds follow a Gaussian distribution with a scale height
of 87 pc,
\citet{mag96} obtain a displacement for the Sun of 14 pc above the mid-plane,
consistent
with analyses based on star counts, Population I objects, dust, and
many other tracers.

\section{Pre-Main-Sequence Star Surveys}

The search for high-latitude pre-main-sequence stars has a
lengthy history primarily based on imaging or objective prism
studies.
These surveys have targeted the H$\alpha$ and \ion{Ca}{I} H\&K emission lines
\citep{her86,sch96}, X-ray emission associated with the magnetospheric
accretion process and the magnetic activity of youth, the infrared excess
due to circumstellar disk thermal emission, and the
nearby stellar associations identified from proper motion studies.
Verification
of candidate PMS stars requires spectroscopy to confirm the signatures
of youth including \ion{Li}{I} absorption, Balmer and other emission lines,
and low surface gravity.

\subsection{H$\alpha$ surveys}

\citet{ste86} surveyed the northern high Galactic latitude sky
$(\delta > -25^{\circ}$ and $|b| > 10^{\circ})$ searching for
candidate H$\alpha$ emission stars using the Burrell Schmidt telescope.
A total of 206 candidates were identified on a series of
$\approx$1300 5.2 $\times$ 5.2 degree red-sensitive objective
prism plates acquired at a dispersion of 1000 {\AA}/mm and a limiting magnitude
$\sim$13.
Subsequent spectroscopic follow-ups
\citep{dow88, wea88,  mah03, emp05} of the resultant catalog [StHA]
have identified several T Tauri stars including a handful at
high Galactic latitude
such as StHA 18 in MBM 12.

\cite{kun92} conducted a deeper objective prism survey (to $V \sim 16$)
with the Schmidt telescope of Konkoly Observatory covering
the MBM 23-44, 49, 50, 53, 54 and 55 molecular
clouds which resulted in 100 candidate H$\alpha$ emission-line stars.
Subsequent spectroscopy at the Isaac Newton Telescope by
\citet{mar96} showed that of the 63 best candidates for PMS status, only
$\sim$20\% were confirmed to have H$\alpha$ in emission and of these
only four are T Tauri stars. These four stars, [K92] 35A, 35B, 37, and 54 are
found in the L134 complex (MBM 33 and MBM 37).

\subsection{IR Surveys}
The IRAS mission produced all-sky imaging at 12, 25, 60, and 100
$\mu$m bands with 10$\sigma$ sensitivities between 0.65 and 3.0
Jy. The Pico dos Dias Survey \citep[PDS;][]{gre92,tor95} utilized the
IRAS Point Source Catalog \citep[PSC;][]{joi85} to search for young
stars including T Tauri stars.  \citet{mag90} used the PSC to search
for low-mass star formation, finding 111 sources projected against 17
translucent and two dark high-latitude ($|b| > 25^{\circ}$) clouds.  A
full search covering $|b| > 30^{\circ}$ using the IRAS Faint Source
Survey \citep[FSS; ][]{mos89} was subsequently performed by
\citet{mag95} yielding 192 candidates.  These surveys targeted objects
with the infrared-excess characteristic of thermal emission from
circumstellar disks following the work of \citet{bei86} and
\citet{mye87}.

\subsection{X-ray Surveys}
Optical (H$\alpha$) and IR surveys are sensitive to properties
associated with Classical T Tauri stars. Since Weak-lined T Tauri stars do not
possess an inner circumstellar disk and thus do not exhibit
either the strong
emission lines due to accretion or an IR excess due to the disk
thermal emission, they are not distinguishable from the field star
population.
WTTS can, however, be selected in X-ray surveys due to strong emission
arising from coronal activity \citep{fei81}.

The Einstein and the ROSAT X-ray satellites were able to
detect significant populations of PMS stars. The Einstein Observatory
\citep{gia79} imaging overlapped 10 of the high-latitude
($|b| > 25^{\circ}$) molecular clouds in the MBM \citep{mag95} catalog:
MBM 7, 12, 18, 20, 27, 30, 36, 43, 54, and 55. Spectroscopy
of the candidate PMS stars in these fields, which are also at
$|b| > 30^{\circ}$, result in only a single
confirmed T Tauri star, which is in MBM 12 \citep{cai95}.
The ROSAT All-Sky Survey \citep[RASS; ][]{tru83} enabled mapping
of the late-type PMS populations around known star formation regions
including Taurus and uncovered previously unknown young stars in the Gould
Belt and the high-latitude field \citep{mag97}.

Other X-ray surveys of high-latitude clouds were performed
by \citet{hea99} and \citet{hea00b} using ROSAT pointed observations and
RASS data. The regions surveyed included the molecular clouds
MBM 7, MBM 12, MBM 40, and MBM 53-55. Optical spectroscopy of candidate
PMS stars confirmed T Tauri stars in MBM 12 and MBM 55.

\subsection{Proper Motion Surveys}

A supplementary technique for identification of PMS populations older
than 10--70 Myr is the search for clustering in the kinematic and spatial
UVWXYZ six-dimensional phase space. When coupled with the ROSAT All-Sky Survey
catalog, this has been an effective means of discovering nearby young
stellar associations \citep[SACY - the Search for Associations Containing
Young Stars Survey; ][]{tor06}. These local populations are discussed
in the chapter by Torres et al. in this Handbook.
Application of proper motion and radial velocity
surveys over the past several years have resulted in the identification
of several nearby stellar groups including TW Hya, $\beta$ Pic, AB Dor,
$\eta$ Cha, $\epsilon$ Cha, Tucana, and Horologium
\citep{zuc04a, lop06}.

\subsection{OB Stars}

\citet{hum47} first identified faint blue stars at high Galactic latitudes.
These stars can be classified as massive Population I stars ejected
from the Galactic disk, as evolved stars such as blue horizontal-branch
[BHB] or post-asymptotic giant branch [PAGB] stars, or as young massive stars
formed in-situ high above the Galactic disk \citep{mar04}. The latter
possibility was subsequently ruled out by \citet{mar06}.
The distinguishing features of stars formed in-situ in the halo are that
they should possess halo kinematics and should have
abundances consistent with that of the intermediate-velocity [IVC] and
high-velocity [HVC] molecular clouds found in this
environment.

\citet{wak01} determined the metallicities and abundances for several
HVCs and IVCs based on spectra from the FUSE satellite and the STIS
and HRS spectrographs on the HST. The derived [Fe/H] values ranged from $-$1 to
0 (nearly solar) suggesting that these clouds have diverse origins
including low-metallicity inflow, possible Galactic fountain clouds, and
tidal remnants, for example from the Magellanic Stream.

[Fe/H] values in selected IVCs and HVCs are $-$0.1 for complex MI,
$-$1.0 for complex CI, $-$0.6 for the Magellanic Stream,
$-$0.3 for the PP Arch, 0.0 (solar) for IV6, IV9, IV19 and the LLIV Arch.
The HVCs appear to be at distances exceeding 5 kpc, while IVCs are
generally within 0.5 to 2 kpc. Typical cloud masses are 0.5 to 8 $\times 10^5$
M$_\odot$ for IVCs and more than $10^6$ M$_\odot$ for HVCs.

Surveys of high-latitude blue stars, e.g. \citet{mar04,mar06},
have tended to rule-out in-situ formation in extraplanar environments
based on metallicity.
As discussed above, near solar metallicities are, however, possible
due to injection into the halo of disk material in fountain and
superbubble flows.
\citet{mar06} combined kinematics and abundance
information for 49 faint high latitude B stars and found
31 Population I runaways, fifteen old
evolved stars (including five BHB stars, three post-HB stars,
a pulsating helium dwarf, and six stars of ambiguous classification),
 one F-dwarf, and two stars that were unclassifiable. \citet{mar06}
concludes that no star in the sample unambiguously shows the
characteristics of a young massive star formed in situ in the halo.

\subsection{SIMBAD Database Search}

The SIMBAD Astronomical Database was searched for
Herbig-Haro Objects [HH] and
the following classes of
stars: confirmed or candidate Young Stellar Objects [Y*O, Y*?],
confirmed or candidate optically detected Pre-Main Sequence Stars
[pr*, pr?],
confirmed or candidate T Tauri stars [TT*, TT?],
Variable Stars of the Orion Type [Or*], and
Variable Stars of the FU Ori Type [FU*] over the entire sky.
The resulting list of 9621 objects included 932 having positions reported as
``No Coord.'' which were subsequently removed from consideration.
The breakdown of SIMBAD object codes in this discarded set were 886 HH, 7 pr*,
25 TT*, 10 Y*?, and 4 Y*O. 107 of the remaining 8689 objects are found at
$|b| \ge 30^\circ$.

Seven objects found in the SIMBAD database were discarded
from this list due to spurious
positions or identifications that resulted in their inclusion in the sample
of high-latitude HH objects and young stars.
The positions of Herbig-Haro objects HH 458,
in Serpens, and HH 26D, in Orion, are incorrectly given in SIMBAD although they
are both correct in the literature, see \citet{dav99} and \citet{dav97},
respectively. Two additional HH objects, HH 20G and HH 23G, discovered by
\citet{gyu78} during their inspection of Palomar Sky Survey plates covering
$|b| < 10^\circ$, had spurious positions reported by SIMBAD.
The massive YSOs [MCB2004b] G41.9 A1 and [MCB2004b] G41.9 A2
\citep{mer04} discovered by the Spitzer Galactic Legacy Infrared Mid-Plane
Survey Extraordinaire (GLIMPSE) Legacy Program are also given incorrect positions by SIMBAD. Finally, although [DG97] 2-1 has an Optical ID of ``YSO faint'' in
Table 1 of \citet{dan97} the text clearly describes this X-ray source
as a candidate AGN.

The results for each class over the entire sky and at high latitudes
are summarized in Table 2. The final column of Table 2 gives the number of
each class whose positions and identifications have been verified and are
Galactic, i.e. not associated with the Large or Small Magellanic Clouds.
For completeness we include six objects classed as young stars
in \citet{hea99} or
in SIMBAD that were subsequently reclassified.
These are Hen 3-1
(aka PDS 1;  RS CVn: K giant plus G dwarf binary), CD-65 150 (aka PDS 3;
a Li-rich K giant), HD 21051 (a K giant, see \citet{smi00}),
V1129 Tau, a G0 eclipsing binary \citep{ote05}, and
GSC 04704-00892 and GSC 05282-02210, both $\sim$100 Myr old K dwarfs
south of Taurus.
Hen 3-1 was the basis for the search that resulted in the
discovery of the $\sim$30 Myr old Tucana-Horologium association even
though it was subsequently understood to not be a member \citep{tor00}.
GSC 04704-00892 and GSC 05282-02210 are examples of X-ray active stars
with weak \ion{Li}{I} absorption that are part of the general young, but
not PMS, population.

\begin{table}[!tb]
\caption{SIMBAD Query Results for Young Stars}
\smallskip
\begin{center}
{
\small
\begin{tabular}{llrrr}
\tableline
\noalign{\smallskip}
Object Type &  Code & All-Sky & $|b| \ge 30^{\circ}$ & Galactic\\
\noalign{\smallskip}
\tableline
\noalign{\smallskip}
Herbig-Haro Object & HH & 2071 & 6 & 2 \\
\tableline
Young Stellar Object & Y*O & 2809 & 13 & 1 \\
PMS Star (optically detected) & pr* & 641 & 30 & 2 \\
T Tauri Star & TT* & 1188 & 58 & 55 \\
Variable Star of the Orion Type & Or* & 729 & 0 & 0\\
Variable Star of the FU Ori Type & FU* & 21 & 0 & 0\\
Candidate Young Stellar Object & Y*? & 38 & 0 & 0 \\
Candidate PMS Star  & pr? & 842 & 0 & 0\\
Candidate T Tauri Star & TT? & 3 & 0 & 0\\
\tableline
Multiple classifications & Y*0,pr*,  & 332 & 0 & 0 \\
 & TT*,Or* & & & \\
\tableline
Star plus Herbig-Haro Object & HH and & 15 & 0 & 0 \\
 & Y*0,TT* & & & \\
\noalign{\smallskip}
\tableline
\end{tabular}
}
\end{center}
\end{table}

The results from SIMBAD were supplemented by 10 stars listed in
\citet{hea99} that had SIMBAD classifications not indicative of
a PMS status. Of these, 3 are classed as components of a Spectroscopic Binary
[SB*], 2 as Emission-Line Stars [Em*], 2 as Variable Stars [V*], 2 as
Stars [*], and 1 as a Peculiar Star [Pe*].
In Table 3 we list the complete set in order of
increasing Galactic latitude with the exception of the six reclassified
stars which are grouped together at the end.
Each table contains the
star name (from SIMBAD), the Galactic coordinates, the spectral
type, and parent molecular cloud or present-day association. We include the
half-dozen H$\alpha$ emission-line stars projected against MBM 18 that
\citet{hea99} attribute to an unpublished paper by Brand et al.
The assignment of molecular cloud or association was based on the
referenced literature. The distinction between the Taurus-Auriga and South
Taurus populations generally falls along lines of Galactic longitude with the
former having $l < 174$, the one exception being the star TYC 664-764-1
identified by \citet{li98}.

\begin{landscape}
\begin{center}
\begin{longtable}{lllllll}
\caption{Young stars at $|b| \ge 30^{\circ}$}
\label{tbl-yso} \\

\hline
\multicolumn{1}{l}{\textbf{Star}} &
\multicolumn{1}{c}{\bf{$l$}} &
\multicolumn{1}{c}{\bf{$b$}} &
\multicolumn{1}{l}{\textbf{SpT}} &
\multicolumn{1}{l}{\textbf{Object}} &
\multicolumn{1}{l}{\textbf{Cloud or}} &
\multicolumn{1}{l}{\textbf{Notes}} \\
\multicolumn{1}{c}{ } &
\multicolumn{1}{c}{\textbf{[deg]}} &
\multicolumn{1}{c}{\textbf{[deg]}} &
\multicolumn{1}{c}{ } &
\multicolumn{1}{l}{\textbf{Type}} &
\multicolumn{1}{l}{\textbf{Association}} &
\multicolumn{1}{c}{ } \\
\hline
\endfirsthead

\multicolumn{7}{c}
{\tablename\ \thetable{} -- continued from previous page} \\
\hline
\multicolumn{1}{l}{\textbf{Star}} &
\multicolumn{1}{c}{\bf{$l$}} &
\multicolumn{1}{c}{\bf{$b$}} &
\multicolumn{1}{l}{\textbf{SpT}} &
\multicolumn{1}{l}{\textbf{Object}} &
\multicolumn{1}{l}{\textbf{Cloud or}} &
\multicolumn{1}{l}{\textbf{Notes}} \\
\multicolumn{1}{c}{ } &
\multicolumn{1}{c}{\textbf{[deg]}} &
\multicolumn{1}{c}{\textbf{[deg]}} &
\multicolumn{1}{c}{ } &
\multicolumn{1}{l}{\textbf{Type}} &
\multicolumn{1}{l}{\textbf{Association}} &
\multicolumn{1}{c}{ } \\
\hline
\endhead

\multicolumn{7}{r}{Continued on next page} \\
\hline
\endfoot

\hline \hline
\multicolumn{7}{l}{Notes:
{\bf 1.} \citet{wei00};
{\bf 2.} \citet{kun92};
{\bf 3.} \citet{mar96};
{\bf 4.} \citet{mag95};
} \\
\multicolumn{7}{l}{{\bf 5.} \citet{li98};
{\bf 6.} \citet{fuh03};
{\bf 7.} \citet{luh01};
{\bf 8.} \citet{mag90};
} \\
\multicolumn{7}{l}{{\bf 9.} \citet{luh04};
{\bf 10.} \citet{hea00a};
{\bf 11.} \citet{gyu82};
{\bf 12.} \citet{li04};
} \\
\multicolumn{7}{l}{{\bf 13.} \citet{fei87};
{\bf 14.} \citet{li00};
{\bf 15.} \citet{smi00};
{\bf 16.} \citet{neu97};
} \\
\multicolumn{7}{l}{{\bf 17.} \citet{neu95};
{\bf 18.} \citet{app98};
{\bf 19.} \citet{zuc04b};
} \\
\multicolumn{7}{l}{{\bf 20.} \citet{fei06};
{\bf 21.} \citet{zuc01};
{\bf 22.} \citet{gre92};
} \\
\multicolumn{7}{l}{{\bf 23.} \citet{rei90};
{\bf 24.} \citet{rei93};
{\bf 25.} \citet{tor00};
} \\
\multicolumn{7}{l}{{\bf 26.} \citet{gre93};
{\bf 27.} \citet{vie03};
{\bf 28.} \citet{wic97};
} \\
\multicolumn{7}{l}{{\bf 29.} \citet{lop06};
{\bf 30.} \citet{zuc08}
}
\endlastfoot
                    HD 141569  &  4.18  &  +36.92  &   B9.5e       &   pr*  & \ldots & 1\\

[K92] 54 & 4.26 & +36.74 & M5.5IV & Em* & MBM 37 & 2,3\\

V* BP Psc &  78.58 & $-$57.22 & mid-F & TT* & \ldots & 4,30\\

2MASS J23101857+1447203 & 89.51 & $-$41.43 & M3V & TT* & MBM 55 & 14\\

                      [LH98] 1  &  155.00  &  $-$35.89  &   G8V         &   TT*  & Tau-Aur & 5\\

                 [FS2003] 0101  &  156.28  &  $-$32.50  &   M:          &   TT*  & Tau-Aur & 5,6\\

                     [LH98] 19  &  156.52  &  $-$32.53  &   K4V         &   TT*  & Tau-Aur & 5\\

                     V* AP Ari  &  156.66  &  $-$30.51  &   K0          &   TT*  & Tau-Aur & 5\\

    [LH98] 3   &  157.69  &  $-$38.03  &   F7V         &   TT*  & Tau-Aur & 5\\

               GSC 01230-01002  &  158.65  &  $-$34.05  &   K6          &   TT*  & MBM 12 & 7\\

                      HD 18831  &  158.74  &  $-$31.24  &   F5          &   TT*  & Tau-Aur & 5\\

                  EM* LkHA 262  &  158.83  &  $-$33.98  &   M1IIIe      &   TT* & MBM 12 & 7,8\\

EM* LkHA 263 & 158.83 & $-$33.98 & M3 & Em* & MBM 12 & 7,8,9\\

                  [HNS2000] 42  &  158.84  &  $-$33.92  &   M5.7...     &   TT* & MBM 12  & 7,10\\

                     [L2001] 8  &  158.86  &  $-$33.31  &   M5.5        &   TT* & MBM 12  & 7\\

                     V* WY Ari  &  158.92  &  $-$33.89  &   K3:... &   TT* & MBM 12  & 7,8\\

                1E 0255.3+2018  &  159.01  &  $-$33.34  &   K3msp       &   TT* & MBM 12 & 4,7\\

                    [L2001] 10  &  159.02  &  $-$33.28  &   M3.2...     &   TT*  & MBM 12 & 7\\

                     [L2001] 9  &  159.27  &  $-$33.64  &   M5.7...     &   TT*  & MBM 12 & 7\\

                  [HNS2000] 53  &  159.38  &  $-$34.07  &   M3          &   Y*O  & MBM 12 & 9,10\\

                     [L2001] 6  &  159.51  &  $-$33.92  &   M5          &   TT*  & MBM 12 & 9,10\\

                    [L2001] 11  &  159.70  &  $-$33.95  &   M5.5        &   TT*  & MBM 12 & 9\\

                TYC 646- 530-1  &  159.96  &  $-$39.95  &   G4V         &   TT*  &  Tau-Aur & 5\\

BD+21 418 & 160.64 & $-$30.03 & G5 & TT* & Subgroup B4 & 29\\

                      [G82b] 3  &  161.38  &  $-$35.93  & \ldots &    HH  & MBM 13 & 11\\

                     [LH98] 20  &  161.57  &  $-$38.16  &   G0V         &   TT*  & Tau-Aur & 5\\

                   EM* StHA 18  &  162.31  &  $-$35.50  &   M3          &   TT*  & MBM 12 & 7,10\\

                      HD 17662  &  163.35  &  $-$41.47  &   G5          &   TT*  & Tau-Aur & 5\\

                     [LH98] 53  &  163.97  &  $-$31.67  & \ldots &   TT*  & Tau-Aur & 12\\

               TYC 1240- 336-1  &  163.97  &  $-$31.67  &   G0V         &   TT*  & Tau-Aur & 5\\

                      HD 18580  &  164.00  &  $-$38.40  &   K2          &   TT*  & Tau-Aur & 5\\

                     [LH98] 40  &  164.79  &  $-$37.19  &   G5IV        &   TT*  & Tau-Aur & 5\\

                TYC 644- 257-1  &  165.18  &  $-$40.31  &   K0          &   TT*  & Tau-Aur & 5\\

               GSC 00644-00256  &  166.31  &  $-$41.74  & \ldots &   TT*  & Tau-Aur & 5\\

                     [LH98] 35  &  166.53  &  $-$40.23  &   K7          &   TT*  & Tau-Aur & 5\\

                         TAP 3  &  166.99  &  $-$33.91  & \ldots &   TT*  & Tau-Aur & 13\\

               TYC 655- 1494-1  &  169.67  &  $-$36.32  &   G2IV        &   TT*  & Tau-Aur & 5\\

 RX J0314.7+1127   &  169.72  &  $-$38.06  &   M1V         &   TT*  & Tau-Aur & 14\\

 RX J0326.3+1131   &  172.29  &  $-$36.11  &   M2V         &   TT*  & Tau-Aur & 14\\

RX J0328.0+1114    &  172.89  &  $-$36.02  &   M3V         &   TT*  & Tau-Aur & 14\\

                     [LH98] 79  &  174.55  &  $-$33.12  &   K0V         &   TT*  & South Tau & 5,16\\

               GSC 00653-00192  &  174.56  &  $-$35.59  &   K3          &   TT* & South Tau & 17\\

                TYC 664- 764-1  &  175.93  &  $-$30.84  &   G7IV        &   TT* & Tau-Aur  & 5\\

               GSC 00060-00489  &  180.24  &  $-$42.71  &   K4   &   pr* & South Tau & 17\\

               GSC 00659-00051  &  180.50  &  $-$31.76  &   K3          &   TT* & South Tau & 17\\

               GSC 00067-01152  &  180.75  &  $-$40.19  &   K7          &   TT* &  South Tau & 17\\

RXJ0344.9+0359 & 183.11 & $-$37.86 & K2 & * & South Tau & 17,18\\

RXJ0354.4+0535 & 183.41 & $-$35.02 & G2 & * & South Tau & 17,18\\

$\sim$6 stars & 189.11 & $-$36.02 & \ldots & H$\alpha$ Em & MBM 18 & Brand et al. [unpub]\\

                  [ATZ98] A123  &  189.30  &  $-$35.14  &   M4V         &   TT*  & South Tau & 18\\

                      HD 25457  &  190.74  &  $-$36.87  &   F5V         &   TT*  & AB Dor & 14,18,19\\

                       GJ 3305  &  198.63  &  $-$30.66  &   M0.5        &   TT*  & $\beta$ Pic & 20,21\\

               GSC 04744-01367  &  203.15  &  $-$30.00  & \ldots &   TT* & \ldots &  22\\

                        HH 123  &  210.82  &  $-$36.61  & \ldots &    HH  & MBM 20 & 23\\

HBC 410 & 210.82 & $-$36.61 &  \ldots & Pe* &  MBM 20 & 8,24\\

V* EW Eri & 210.86 & $-$36.56 & K7IV & V* & MBM 20 & 8,24\\

                     BD-15 808  &  211.86  &  $-$37.51  &   G4V         &   TT*  & MBM 20 & 14\\

HD98800Aa & 278.40 & +33.80 & K5V & SB* & TW Hydrae & 22\\

HD98800Ba & 278.40 & +33.80 & K7V & SB* & TW Hydrae & 22\\

HD98800Bb & 278.40 & +33.80 & M1V & SB* & TW Hydrae & 22\\

                TYC 8474-24-1  &  293.77  &  $-$64.10  &   F2          &   TT* & \ldots & 22,27  \\

       2MASS J14104963-2355290  &  325.33  &  +35.47  &   K2          &   TT* & Gould Belt & 28\\

                     HD 125340  &  327.80  &  +35.22  &   G6IV      &   TT* & Gould Belt & 28\\

               TYC 6138-383-1  &  329.46  &  +42.17  &   G9          &   TT* & Gould Belt & 28\\

                      [K92] 35a  &  357.84  &  +37.52  &  M4V           &   TT* & MBM 33  & 2,3\\

            [K92] 35b  &  357.84  &  +37.52  &  M5IV           &   TT* & MBM 33 &  2,3\\

                      [K92] 37  &  358.29  &  +37.78  &   M4.5IV      &   TT* &  MBM 33 & 2,3 \\

\hline
\hline
\multicolumn{7}{l}{Reclassified PMS candidates} \\
\hline
\hline
                      HD 21051  &  170.87  &  $-$35.66  &   K0III-IV    &   TT*  & Tau-Aur & 5,14,15 [K giant]\\

               GSC 04704-00892  &  176.98  &  $-$56.76  &   K4          &   TT* & South Tau & 17 [$\sim$100 Myr ZAMS] \\

V* V1129 Tau & 177.40 & $-$32.32 & G0 & TT* & South Tau & 5 [Eclipsing binary]\\

               GSC 05282-02210  &  178.40  &  $-$63.33  &   K0          &   TT* & South Tau & 17 [$\sim$100 Myr ZAMS] \\

Hen 3-1 & 280.58 & $-$59.07 & dKe? & V* & Horologium & 22,25 [RS CVn]\\

                     CD-65 150  &  283.37  &  $-$46.19  &   K1III       &   TT*  & \ldots & 22,26 [Li-rich K giant]\\
\end{longtable}
\end{center}
\end{landscape}

\begin{landscape}
\begin{figure}
\begin{center}
\includegraphics{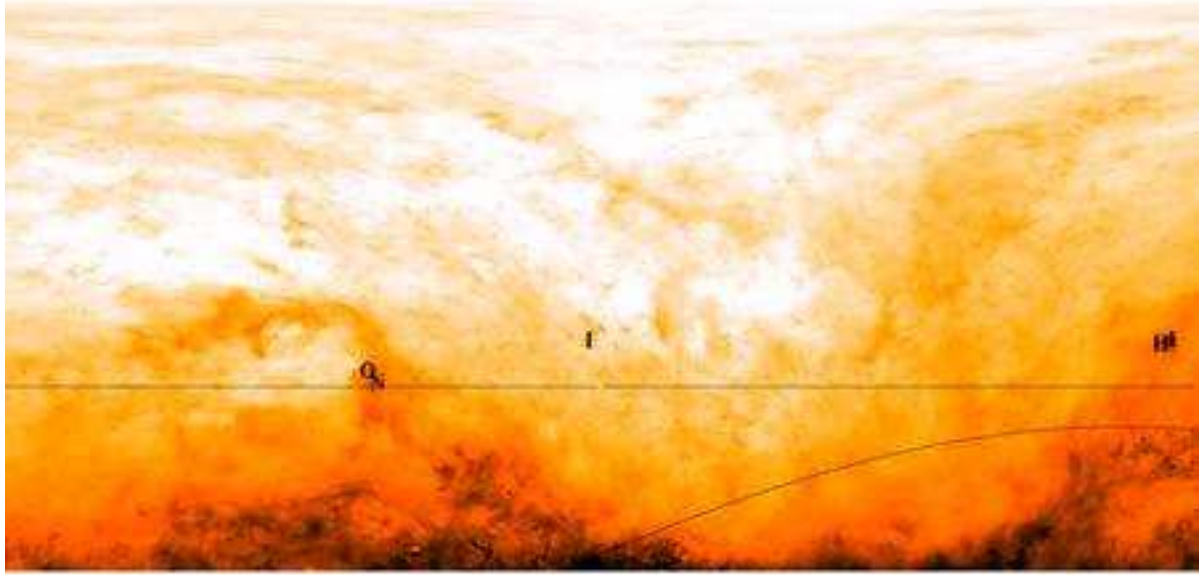}
\end{center}
\caption{High-latitude star formation regions and young stars in the
Northern First and Second Galactic Quadrants $(0^{\circ} < l <
180^{\circ}; b > 0^{\circ})$.  This region contains the HD 141569
system {\bf (1)} consisting of a Herbig AeBe star and two M stars.
The molecular clouds shown are MBM 37 {\bf (H)} and MBM 41-44 {\bf
(I)}. Within the Polaris Flare are two high-latitude translucent
clouds IREC 165 {\bf (N)} and L1311 {\bf (O)}.
The background is from the dust map of Schlegel et al. (1998).}
\label{fig-gal12N}
\end{figure}
\end{landscape}

\begin{landscape}
\begin{figure}
\begin{center}
\includegraphics{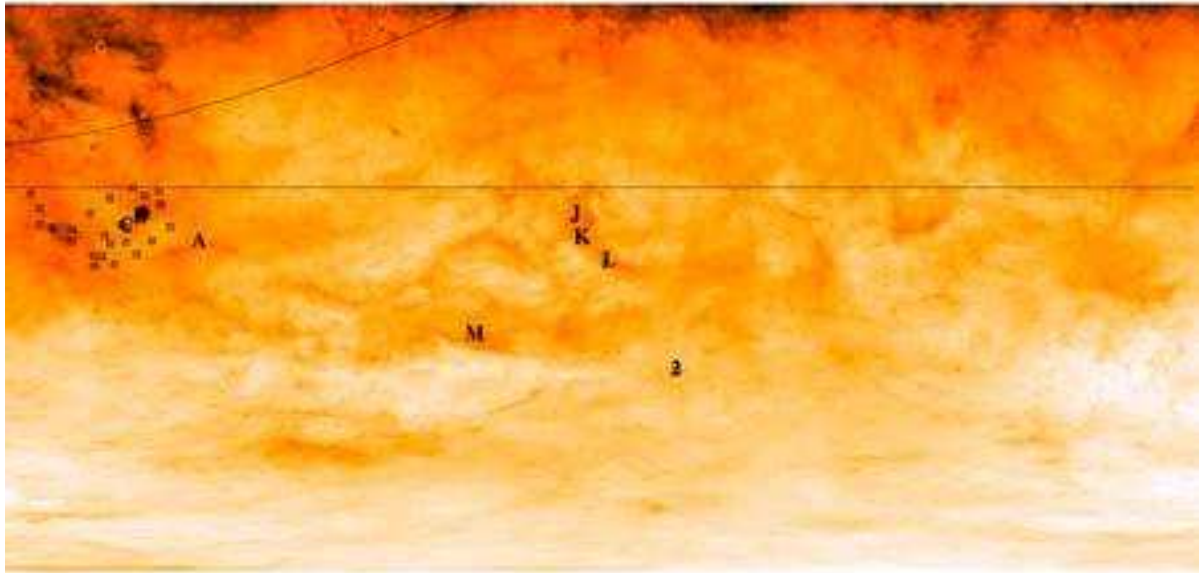}
\end{center}
\caption{High-latitude star formation regions and young stars in the Southern
First and Second Galactic Quadrants
$(0^{\circ} < l < 180^{\circ}; b < 0^{\circ})$.
A second isolated Herbig AeBe star, BP Psc {\bf (2}), is found in this
region. Six molecular clouds - MBM 7 {\bf (A}), MBM 12 {\bf (B)},
MBM 13 {\bf (C)}, MBM 53 {\bf (J)}, MBM 54 {\bf (K)} and MBM 55 {\bf (L)}
are identified here.
MBM 12 is one of two active high-latitude star formation regions.
The southernmost cloud is the translucent cloud IREC 146 {\bf (M)}
which is part of a high-latitude CO shell identified by
\citet{cha06}.
The PMS stars within the South Taurus region are
evident in the left-hand portion of this map.
The background is from the dust map of Schlegel et al. (1998).
\label{fig-gal12S}}
\end{figure}
\end{landscape}

\begin{landscape}
\begin{figure}
\begin{center}
\includegraphics{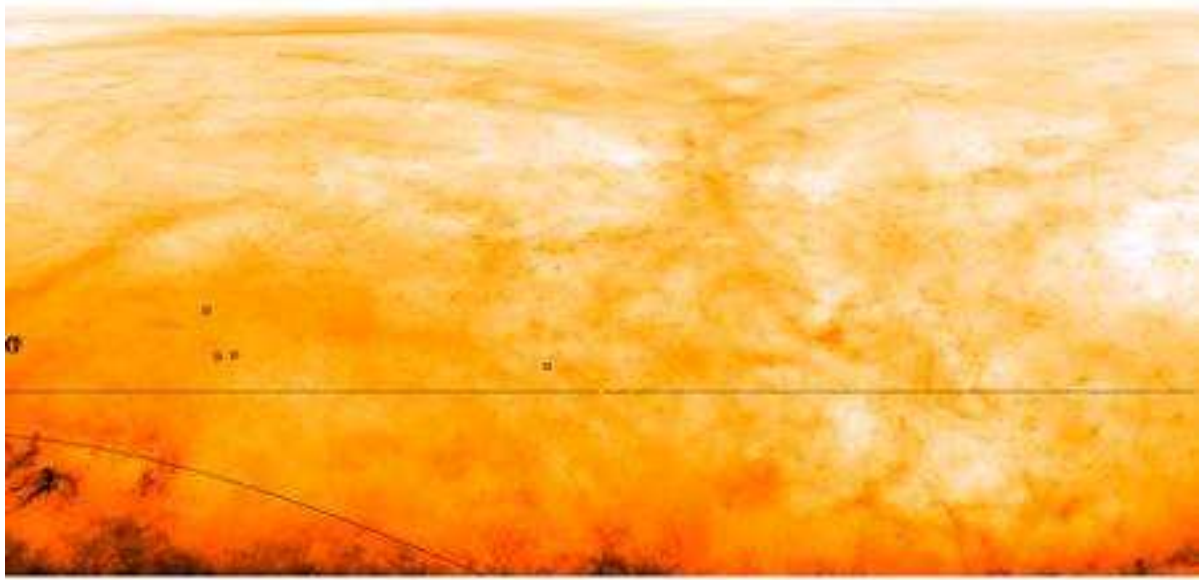}
\end{center}
\caption{High-latitude star formation regions and young stars in the Northern
Third and Fourth Galactic Quadrants
$(180^{\circ} < l < 360^{\circ}; b > 0^{\circ})$.
With the exception of MBM 33 {\bf (G)}, a few Gould Belt members, and
HD 98800 (TWA 4), this appears to be relatively devoid of recent star
formation.
The background is from the dust map of Schlegel et al. (1998).
\label{fig-gal34N}}
\end{figure}
\end{landscape}

\begin{landscape}
\begin{figure}
\begin{center}
\includegraphics{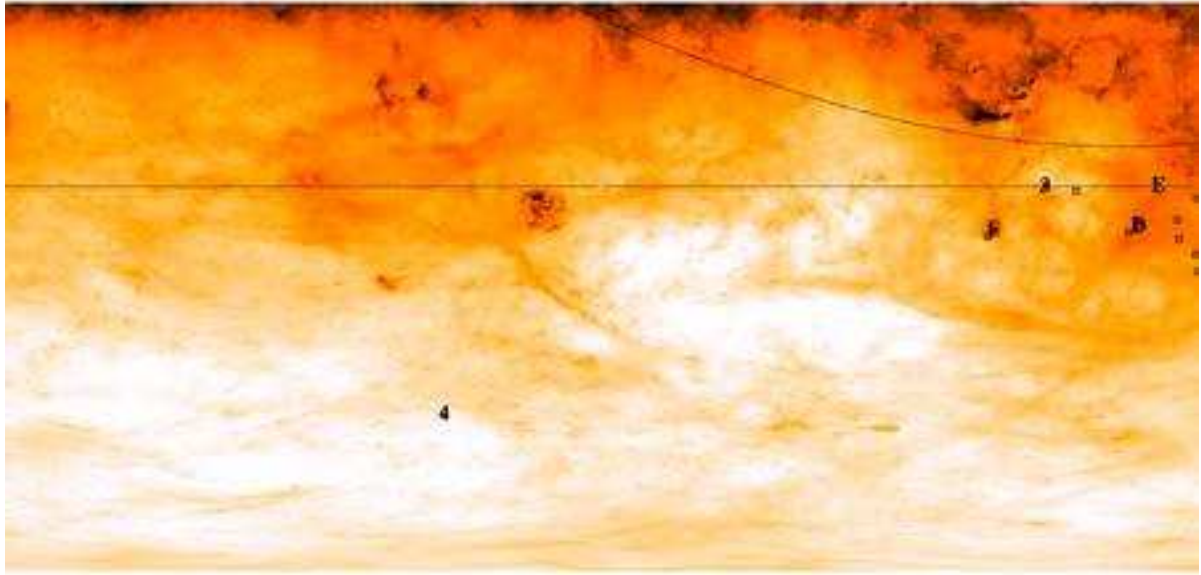}
\end{center}
\caption{High-latitude star formation regions and young stars in the Southern
Third and Fourth Galactic Quadrants
$(180^{\circ} < l < 360^{\circ}; b < 0^{\circ})$.
The isolated PMS stars in this region are
GSC 04744-01367 {\bf (3)} and TYC 8474-24-1 {\bf(4)},
of which the latter is a Herbig AeBe star. The molecular clouds
seen just south of the Taurus and Orion star formation regions are
MBM 18 {\bf (D)}, MBM 19 {\bf (E)}, and MBM 20 {\bf (F)}.
MBM 18 and MBM 20 contain the Lynds dark clouds L1569 and L1642.
L1642 is one of the Orion outlying clouds and is an active star formation
region.
The background is from the dust map of Schlegel et al. (1998).
\label{fig-gal34S}}
\end{figure}
\end{landscape}

\section{The High-latitude Star Formation Regions}

Pre-main-sequence stars at high Galactic latitude are found as
dispersed populations in large-scale structures, within
individual molecular clouds,
stellar associations, and in isolation. For each region we
discuss the distance, age, massive and low-mass stellar populations,
and the atomic and molecular gas.
Figures \ref{fig-gal12N} through \ref{fig-gal34S}
identify the specific molecular clouds and young
stars using the \citet{sch98} extinction maps as backdrops.
The background for each image is the \citet{sch98} FIR-derived
dust map displayed on a log scale ranging from $E(B-V) = 0.01$ to
$E(B-V) = 3.23$ $(A_V = 10)$.
Molecular clouds discussed in the text and
the three translucent clouds distant from the Gould Belt are
marked by letters. The four isolated high-latitude PMS stars are
labeled by numbers.
The Gould Belt mid-plane and the $|b| = 30^{\circ}$ definition for
Galactic high-latitude are shown for reference.

\subsection{Large-scale Structures}

\subsubsection{The Gould Belt:}

\citet{gou79} noted that the
brightest stars on the sky were
aligned along a great circle crossing the Milky Way at an angle close to
20$^\circ$. This structure, known as the Gould Belt, has been the subject of
extensive studies over the intervening century.

The Gould Belt appears on the sky as a band approximately
20$^{\circ}$ wide
having an ascending node $l_\Omega = 282^{\circ}$ and an
inclination $i = 24^{\circ}$, thus the Gould Belt populations
appear at $b > 30^{\circ}$ and
$b < -30^{\circ}$ around $l = 12^{\circ}$ and $l = 192^{\circ}$,
respectively.
The Gould Belt is described
as an ellipsoidal shaped ring with semi-major and
semi-minor axes of 500 pc and 340 pc \citep{gui98}. The center of this
structure is approximately 200 pc distant from the Sun towards
$l = 130^{\circ}$. The composition of the Gould Belt includes interstellar
matter and a combination of Population I disk stars of ages
up to 80 Myr, OB associations, and young clusters. Proposed
formation scenarios for the Gould Belt include
collisions of high velocity clouds with the
Galactic disk \citep{com92, com94} or induced star formation from winds
originating in supernovae or the central Cas-Tau association \citep{bla91}.
\citet{har01} suggest that the Gould Belt is not truly a coherent
structure but is due to the motion of interstellar material
around and out of the Galactic plane driven by complex and interacting
flows.

\citet{wic97} identified three high-latitude Gould Belt stars
near $l = 327^{\circ}$ in their spectroscopic survey
spanning $b = -5^{\circ}$ to $50^{\circ}$ of RASS selected T Tauri
candidates in Lupus. Li-rich RASS sources in other regions of the sky,
including South Taurus, may be also associated with the Gould Belt.

\subsubsection{The Taurus-Auriga Complex:}

The southern high-latitude skies between $l = 150^{\circ}$ and
$l = 200^{\circ}$ are dominated by the dispersed stellar populations
associated with the young ($<$ 10 Myr) and nearby (140 pc)
Taurus-Auriga star formation region (see the chapter by
Kenyon et al. in this Handbook). The older
Gould Belt populations in this region are 300 to 500 pc distant.

Spectroscopy of optical counterparts of RASS sources south of Taurus
by \citet{mag97} and \citet{mar99} shows that a significant fraction
of the RASS-selected stars are not young WTTS \citep{neu95} but,
on the basis of their Li abundance, are older post T Tauri stars [PTTS].
The presence of PTTS in the central regions of Taurus suggests that
low levels of ongoing star formation have been occurring there for
over 10 Myr. The semi-regular pulsating G0 star V1129 Tau was identified
as a possible older member of Taurus-Auriga by \citet{li98} on the
basis of significant but weak \ion{Li}{I} absorption. V1229 Tau is also
classed as an eclipsing binary by \citet{ote05}.

Variability studies comparing TTS in South Taurus, Taurus-Auriga, and
MBM 12 indicate no sigificant difference in the distribution of
rotational periods \citep{bro06}. Given the limited sample sizes
this study was unable to distinguish between the ejection and small
cloudlet formation scenarios for the dispersed young stars. If disk braking
plays a major role in the angular momentum evolution of young stars, then
due to the truncation of the circumstellar disk by tidal forces the
dispersed population should be rotating more rapidly. \citet{bro06} find
periods of 4.7 and 1.34 days for two South Taurus TTS
GSC 00653-00192 and GSC 00060-00489,
respectively. RXJ0344.9+0359 is clearly variable but no period could be
determined. The rotational periods of [LH98] 20 and [LH98] 53 have been
found to be 1.0 and 0.7 days \citep{xin07a,xin07b}.

\subsection{Molecular Clouds}

In this section we discuss the individual high latitude clouds that
either contain spectroscopically confirmed
pre-main-sequence stars or have been suspected of harboring star formation.
We use the MBM catalog designations based on the CO (J=1-0) survey
of \citet{mag85} and the literature review of \citet{mag96}.

\subsubsection{MBM 7:}

\citet{hea99} conducted spectroscopic followup of ROSAT
All-Sky Survey targets projected against MBM 7 $(l=150.47, b=-38.07)$.
In their survey of X-ray active stars towards MBM 7 and MBM 55 (see below)
\citet{hea99} conclude that the majority, which are both Li-weak and near
to the main seqeuence, are of the general nearby stellar population
younger than 150 Myr.
The estimated distance to MBM 7 is 75 to 175 pc. This
cloud is southwest of the Taurus-Auriga complex and within a large
arc of dust and gas that has been swept up by either supernovae
or stellar winds. MBM 7 is of interest as a laboratory for triggered
star formation processes.

\begin{figure}[ht]
\plottwo{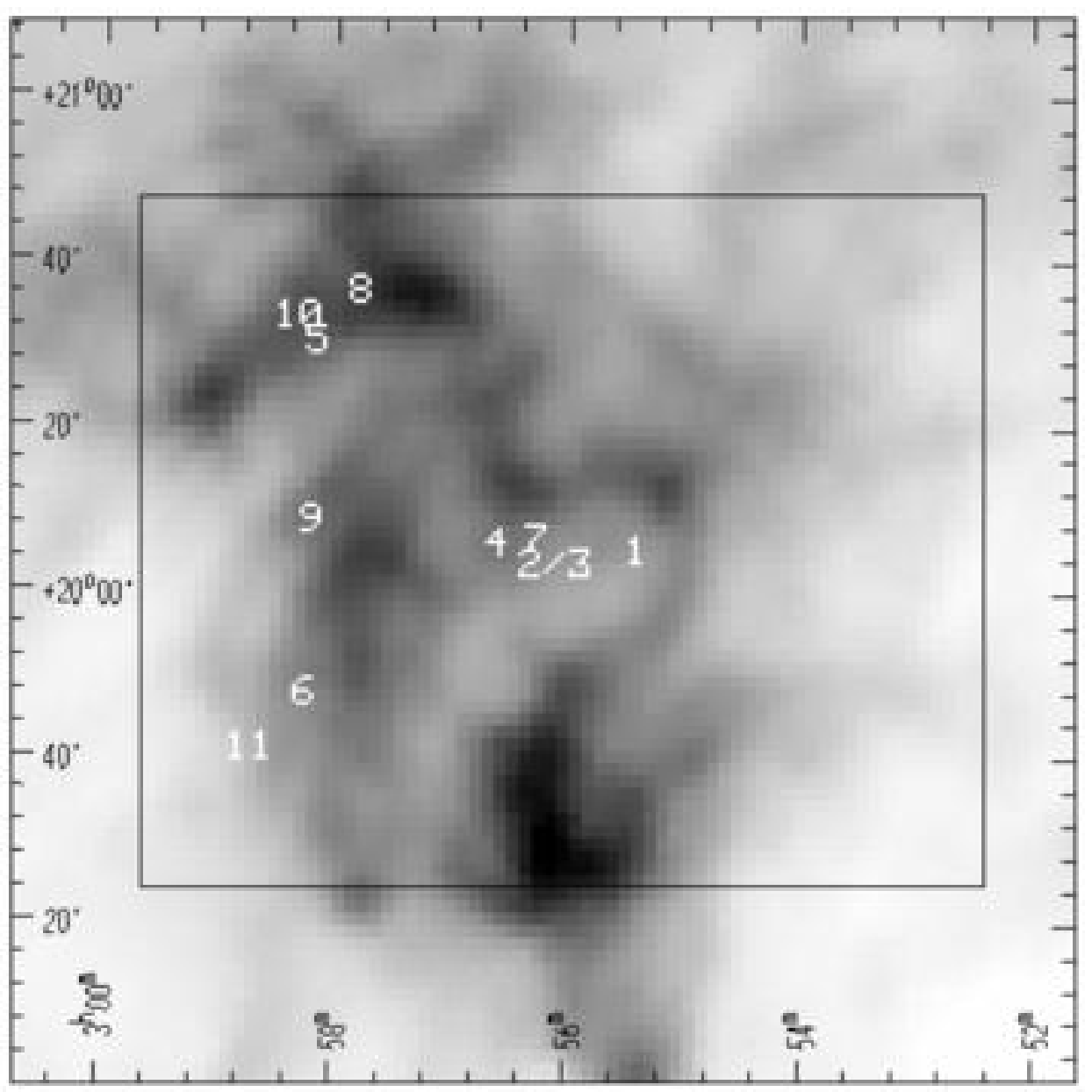}{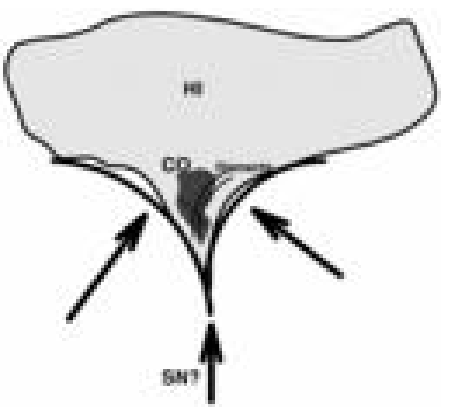}
\caption{The left panel shows the locations of MBM 12A 1--11 marked on an
IRAS 100 $\mu$m image of the MBM 12 cloud \citep[from][]{luh01}.
On the right
 \citep[from][]{mor97}
is a depiction of a possible triggered
star formation scenario
due to an external supernova.
\label{fig-mbm12a}}
\end{figure}

\subsubsection{MBM 12/MBM 13:}

The MBM 12 cloud $(l=159.35, b=-34.32)$ is an active region of
low mass formation. MBM 12 includes L1457
which is the only dark cloud as defined by $A_V > 5$
found at high Galactic latitudes.
\citet{hea00b} obtain a distance estimate of $58\pm5 < d < 90\pm12$
based on the presence and absence of \ion{Na}{I} D line absorption in
stars with Hipparcos parallaxes, which led to the assertion that MBM 12
was the nearest molecular cloud and located well within the
Local Bubble of hot ionized gas.
Recent studies by \citet{luh01} and \citet{and02}
now suggest that the cloud is at 275 pc.
\citet{luh01} found extinction at 65, 140, and 275 pc
along the line of sight with the 275 pc distance giving
the most plausible location on the H-R diagram. A subsequent analysis
by \citet{and02} find a supporting result of $360\pm30$ pc with
a foreground, less opaque, layer of extinction at $\sim$80 pc.
\citet{str02} derived a distance estimate of 325 pc based on photometry
conducted in the Vilnius seven-color system for 152 stars.
T Tauri candidates in MBM 12 have been selected by
H$\alpha$ and X-ray surveys, with spectroscopic followup revealing
a population of a dozen low-mass PMS stars
\citep{ste86,her88,hea00b,luh01}. The latter survey is complete to
0.03 M$_\odot$ (H$\sim$15) with the verified members having
masses down to 0.1 M$_\odot$. The age of the MBM 12 association
is estimated as $2^{+3}_{-1}$ Myr on the basis of relative
\ion{Li}{I} strengths and on H-R diagrams \citep{luh01}.

Figure \ref{fig-mbm12a} depicts the IRAS 100 $\mu$m image of the
MBM 12 cloud with the positions of
the young stars (MBM 12A 1--12).
MBM 12A 2 and 3 (LkH$\alpha$ 262 and 263)
are a wide binary system.
MBM 12A 12 (StHA 18) is projected against the MBM 13 molecular cloud,
2.5$^{\circ}$ southeast of MBM 12.
MBM 12A 12 is a likely member of this association.

\begin{figure}
\plotone{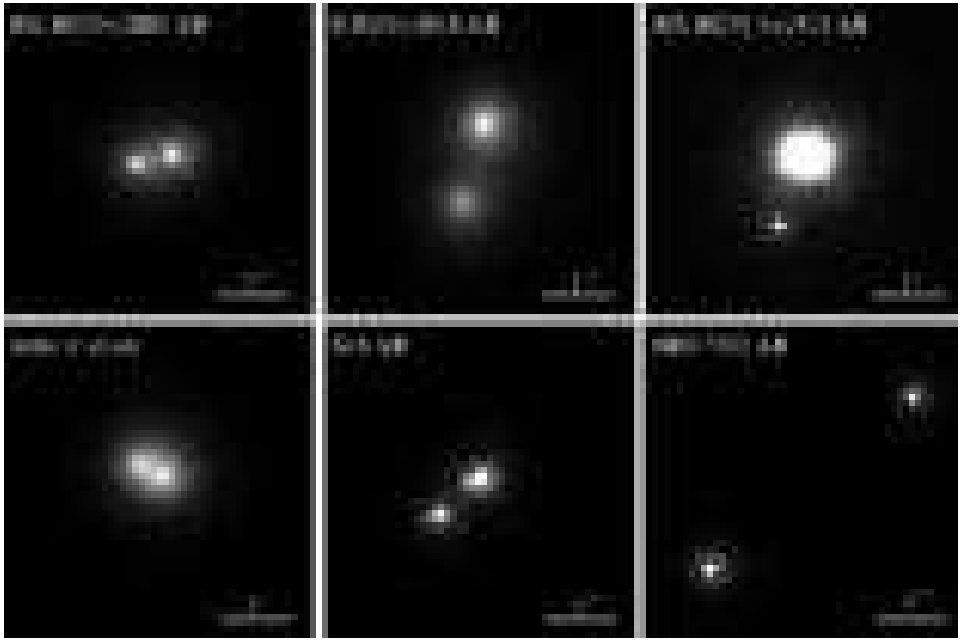}
\caption{CFHT AO images of binaries in MBM 12 acquired by
\citet{cha02}.
\label{fig-mbm12b}}
\end{figure}

\citet{cha02} obtained adaptive optics observations with the
Canada-France-Hawaii Telescope discovering
six binaries,
LkH$\alpha$ 264, E 0255+2018, RX\linebreak J0255.4+2005, StHA 18, MBM 12A 10,
RX J0255.3+1915, and confirming the binary nature of HD 17332 (see
Figure \ref{fig-mbm12b}).
In addition they detected a possible quadruple system
composed of the close binary LkH$\alpha$ 263AB (separation of
$\sim$0.41$''$), LkH$\alpha$ 262 located $\sim$15.3$''$ from
 LkH$\alpha$ 263A,
and of LkH$\alpha$ 263C, located $\sim$4.1$''$ from LkH$\alpha$ 263A.
The latter is a nebulous object interpreted as a
disk oriented almost perfectly edge-on and seen in scattered light.
These results suggest that the binary fraction in MBM 12 is high compared to the
field and to IC 348.

\citet{bro06} report periodic variability in four MBM 12 TTS
attributed to
rotational modulation of photospheric features. The inferred rotational
period ranged from 3.36 to 7.4 days, although for the CTTS
LkH$\alpha$ 264 the observed period differed from results obtained during
previous epochs suggesting a more complex mechanism for variability.
The WTTS 1E 0225.3+2018 showed
strong variability, but poor sampling in this study
rendered determination of an accurate period impossible. \citet{pin06}
find the computed mass accretion rates based on $U$-band veiling
for GSC 01230-01002 (RXJ0255.4+2005), LkH$\alpha$ 264, and 1E 02553+2018
to be in the range 10$^{-8}$ to 10$^{-7}$ M$_{\odot}$ yr$^{-1}$, consistent
with results from other star formation regions of similar age.

Millimeter continuum observations of TTS in MBM 12 reveal that
the disk masses are less than 0.1 M$_\odot$.
These mass limits are consistent with results in Taurus-Auriga
and $\rho$ Oph, however massive disks as found around T Tau, DG Tau, and
GG Tau can be ruled out.
The 3 mm dust continuum survey by \citet{hog02}
around eight TTS in MBM 12 (6 in 5 fields;
2 from literature) limits $M_{disk} <$ 0.04--0.09 $M_{\odot}$ assuming
a gas-to-dust ratio of 100 and a distance of 275 pc. Coadding of fields
places upper limits on the average disk mass to 0.03 $M_\odot$.
A subsequent survey by \citet{ito03} detected, at the 3$\sigma$ level and
above, emission at 2.1 mm from three MBM 12 TTS. The derived disk masses
are 0.048 M$_{\odot}$ for LkH$\alpha$ 262, 0.085 M$_{\odot}$ for
LkH$\alpha$ 264, and 0.071 M$_{\odot}$ StHA 18.
Based on 450~$\mu$m and 850~$\mu$m studies, \citet{hog03}
assign dust opacity model dependent masses ranging between
0.02 to 0.2 M$_{\odot}$ for LkH$\alpha$ 262 and LkH$\alpha$ 264A.
For LkH$\alpha$ 263ABC and StHA 18 \citet{hog03}
estimate masses that are an order of magnitude lower.

\citet{hon06} find that in a study of 30 TTS, the
silicate dust around WY Ari (LkH$\alpha$ 264)
has anomalously low large component and
crystallinity fractions given the strength of the 10~$\mu$m
silicate feature. The dominant component, as inferred from
model fits to $N$-band spectra, appears to be submicron olivine grains.
\citet{hon06} speculate that accretion-induced turbulence may allow
the smaller grains to drift higher thus making the disk surface rich
in small grains.

In their millimeter wave survey [$^{12}$CO(1-0),
$^{13}$CO(1-0), and CS(2-1)] of high-latitude
clouds for proto-brown dwarfs, \citet{pou93} found no clearly self-gravitating
low mass clumps in MBM 26, MBM 27--29, MBM 30, MBM 31--32, and MBM 41--44. They
did, however, identify two objects in MBM 12 whose masses were within a
factor of 3 of being gravitationally bound. One of these two clumps,
MBM 12-1, has a dense core detected in CS(2-1) emission and is
apparently in hydrostatic equilibrium.


\subsubsection{MBM 16:}

MBM 16 ($l=170.60, b=-37.27$) is a large translucent cloud located
$\sim 20^{\circ}$ south of the Taurus clouds. On the basis of interstellar
absorption, \citet{hob88} assigned a distance of 60 to 85 pc.
\citet{mag03} adopt a distance of 100 pc, citing arguments concerning
X-ray shadowing that MBM 16 is within the boundary of the Local Bubble
\citep{kun97,cox87}. Four of the WTTS candidates identified by \citet{li00}
are projected against MBM 16.

\subsubsection{MBM 18/MBM 19:}

The MBM 18 ($l=189.10, b=-36.02$) and MBM 19
$(l=186.03, b=-29.93)$ clouds have been studied by Brand \& Wouterloot
\citep[unpublished, see][]{hea99}
who identified candidate T Tauri stars on the basis of
H$\alpha$ emission and by Larson and collaborators
\citep{lar03,cha04}.
MBM 19 extends north from the MBM 18 high-latitude
cloud towards the Taurus SFR and is a suspected site of low-mass
star formation.
Although MBM 19 exhibits low, relatively smooth
100 $\mu$m emission and modest $A_V$, the $^{12}$CO(1-0) in
MBM 19 shows clumpy emission with line intensities $>$ 3K. \citet{cha04}
find $\sim$20 stars within MBM 18 and MBM 19
with possible circumstellar disk signatures on the
basis of 2MASS colors.
\citet{cha04} determine a distance to MBM 18 (the Lynds dark cloud L1569)
of 80$\pm$20 pc based
on line-of-sight interstellar reddening. They also see only
slight evidence for an increase of $R_V$ with increasing $A_V$ suggesting
that dust grain growth has not occurred in the denser regions of the
molecular cloud.

\subsubsection{MBM 20:}

MBM 20 ($l=210.89, b=-36.51$) is a member of the Orion outlying
clouds discussed in the chapter by Alcala, Covino, and Leccia
in this Handbook. The distance to MBM 20 has been measured as
between 100 to 160 pc; significantly closer than the Orion OB1
association (330 to 450 pc).
\citet{hea00b} obtain a distance estimate for MBM 20
of $112\pm15 < d < 151\pm21$ on the basis of \ion{Na}{I} D line
absorption studies of Hipparcos stars. \citet{leh07}, based on ISO
200~$\mu$m mapping of the L1642 Lynds dark cloud in MBM 20,
find evidence for a two-fold increase in the dust emissivity
over the temperature range 19 K to 14 K. They attribute this to
an increase of the dust absorption cross-section, perhaps due
to coagulation of dust grains within the cold molecular core.

Two binary CTTS, L1642-1 (aka HBC 413 and EW Eri) and
L1642-2 (aka HBC 410), were
identified by \citet{san87}. Evidence for a bipolar outflow is detected
in $^{12}$CO(1-0) observations near L1642-2 \citep{lil89}.
The X-ray active G4V star BD-15 808 identified by \citet{li00} is projected
within $\sim$1 degree of MBM 20. \citet{car05} assign an age of 30 to 100 Myr
to BD-15 808 and cite a distance of 140 pc, consistent with that of MBM 20.
This star was observed during the
Formation and Evolution of Planetary Systems (FEPS) {\it Spitzer}
Legacy program and is likely an evolved system based on the lack of
an IR excess in the IRAC bands \citep{sil06} and little or no emission
at 1.2~mm \citep{car05}. BD-15 808 may be a member of a general $<$ 150 Myr
population noted by \citet{hea99} and not be associated with MBM 20.

\subsubsection{MBM 33/MBM 37:}

MBM 33 $(l=359.06, b=+36.75)$ and MBM 37 $(l=5.70, b=+36.22)$
are cometary globules
within the L134
complex located at the northernmost extension of the Sco-Oph
star formation region.
Early studies of MBM 33 included spectroscopy and
multi-band photoelectric observations of the L1778 dark cloud
by \citet{lyn65}.
\citet{fra89} derived a distance of 110$\pm$10 pc
on the basis of color excess observed in SAO stars of clouds within the
region $l = 359^{\circ}$ to 12$^{\circ}$ and $b = 35^{\circ}$ to
39$^{\circ}$ which includes the dark clouds
L134, L169, and L1780.
Furthermore, \citet{fra89} suggests that these clouds are immersed in a
sheet-like absorbing structure related to the surface of the Loop
I Bubble.

Two candidate WTTS (K35AB and K37) are identified
in projection against
MBM 33 \citep{mar96}.
These stars have a mid-M spectral type, H$\alpha$ EW of $-$4 to $-$9{\AA},
and \ion{Li}{I} EW of 600 to 700 m{\AA}. An additional WTTS, K54, is seen
near MBM 37 having a mid to
late M spectral type,  H$\alpha$ EW of $-$15{\AA},
and \ion{Li}{I} EW of 800 m{\AA}.

\citet{rid06} find that the L1780 cloud within MBM 33 contains a
dense core having a temperature between 14 K and 15 K, assuming a dust
emissivity index
$\beta$ = 2 and based on IRAS and ISO observations spanning 12~$\mu$m to
200~$\mu$m. There appear to be separate cold and warm dust components
as the spatial distribution of the 12 $\mu$m, 25 $\mu$m, and 60 $\mu$m
emission differs from the longer wavelength data that trace the core.
The $^{13}$CO core of L1780 appears to be
in virial equilibrium, but starless, as no candidate young stars are
seen in IRAS, ISO, and 2MASS imaging data.

L183 (=L134N), a quiescent cloud lacking signs of protostellar-driven
outflow activity, has been a popular subject for astrochemists starting
with the H$_2$CO study by \citet{eva76}. The L134N designation,
first used by \citet{eva76},
arises from \citet{lyn62} grouping L134, L169, and L183 as
different sections of a single object (ID=85).
Unlike the dark cloud L1448, which contains the extremely young
outflow L1448-mm, L183 is deficit in SiO and complex organic molecules
(COMs)
such as CH$_3$OH. \citet{req07} propose that the observed COM
abundances are consistent with a ``universal'' grain mantle
composition that is locally changed by the process of low-mass star formation.

Spectroscopic studies targeting molecular gas phases
become difficult in cold cores due to the depletion of molecules
onto ice mantles.
 N$_2$H$^+$ and N$_2$D$^+$ observations of L183 by \citet{pag07} indicate
that the gas is thermalized with the $\sim$7 K dust and that
 N$_2$H$^+$ is significantly depleted starting at densities
5--7 $\times 10^5$ cm$^{-3}$. \citet{pag07} suggest that the relatively
less depleted N$_2$D$^+$ is a viable tool for study of cold cores.
\citet{aky07} studied the spatial variation of NO abundance in L183
and L1544, another pre-protostellar core, finding evidence for
depletion towards the density peaks of each. In the case of L183 the
central minimum in the fractional abundance of NO is shifted by
10 arcsec in R.A., or 1000 AU, relative to the dust peak.

\begin{figure}
\centering
\includegraphics[draft=False,width=\textwidth]{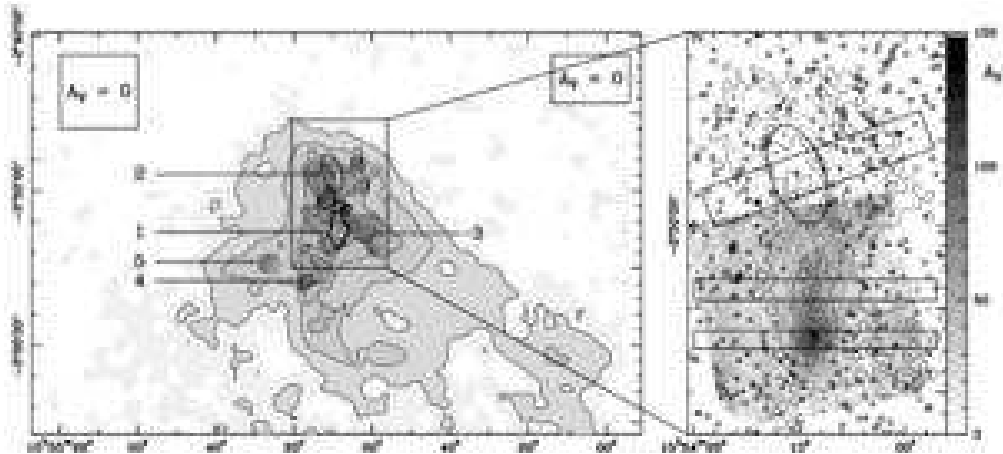}
\caption{Composite extinction image of L183 (=L134N) from \citet{pag04}.
The numbered peaks are those with molecular counterparts. Peaks 1 and 2
are candidate prestellar cold cores with dust temperatures $\sim$7 K.
The stripes in the right-hand panel indicate regions used to study the
cloud morphology. Background stars detected in $H$ and $K'$ and
used to estimate extinction are marked by stars and asterisks.
\label{fig-l183}}
\end{figure}

L183 is associated with MBM 37 and contains five extinction peaks
shown in Figure \ref{fig-l183} that related to known molecular peaks.
Of these, two are judged by \citet{pag04} to be candidate prestellar
cores, of which peak 1 is thought to be more highly evolved due to its
logotropic profile \citep{mcl97}.  The total mass of the core at Peak
1 within a radius of 1 arcmin was determined by \citet{pag04} to be
$\sim$2.5 M$_\odot$.  L183 is designated as a ``bright'' core by
\citet{kir05} who measured integrated flux densities within a
2.5-arcmin diameter aperture of 12.8 Jy at 450 $\mu$m and 4.7 Jy 850
$\mu$m determining a total mass of 1.8 M$_{\odot}$. L183 exhibits
different morphologies at these two submillimeter bands with only the
denser southern core (Peak 1) being detected at 850 $\mu$m.

\subsubsection{MBM 41--44:}

The MBM 41--44 complex, also known as the Draco Nebula,
is an intensely studied intermediate velocity cloud
v$_{lsr}$ = $-$21 km/s.  This object was discovered on the basis
of a \ion{H}{I} survey conducted by \citet{goe83}.
A detailed \ion{H}{I} map of the Spitzer First Look
Survey by the 100 m Green Bank Telescope \citep{loc05} detected a portion
of the Draco nebula at a peak \ion{H}{I} column density of
$7.8 \times 10^{19}$ cm$^{-2}$.
The cloud subtends 5
degrees on the sky and exhibits
a number of compact cores and a distinct filamentary
structure trailing towards high-$b$ and high-$l$ from
the high-density complex at $(l \sim 90, b \sim 38)$.
Distance estimates vary with those based on
\ion{Na}{I} D absorption suggesting $463^{+192}_{-136}$ to $618^{+243}_{-174}$ pc
\citep{gla98}.
Star count analyzes and reddening studies have resulted in significantly
greater distance estimates including 800 to 2500 pc \citep{goe86}, and
800 to 1300 pc \citep{pen00}.

CO 2.6 and 2.7 mm and H$_2$CO (6.2 mm) studies reveal two
molecular clumps which, if the nebula is more than 800 pc distant, have
masses of 27 and
219 M$_\odot$ \citep{meb85}. On the basis of virial theorem
arguments these molecular clumps would be gravitationally bound for
a distance to the Draco IVC exceeding 600 pc.
\citet{joh86} and \citet{joh87} examine optical
counterparts to IRAS sources projected against the nebula suggesting that
33 of the optically unidentifiable sources may be similar to Bok globules
although many of these may be sharp gradients in the 100 $\mu$m cirrus
misidentified by the PSC source detection algorithm.

On the basis of the plume morphology seen in the IRAS
100 $\mu$m images, \citet{ode87} propose that the interaction of the Draco
IVC with the Galactic plane is a subsonic, hydrodynamic phenomenon with a
Reynolds number between $\sim$10 and $\sim$20.  If this
interpretation is correct then we should not expect the infall of the
cloud to be effective in triggering star formation along the leading
edge. As the cloud is not self-gravitating, turbulent compression is
needed to establish a scale hierarchy and subsequent collapse \citep{mac04}.
Since the formation of turbulence requires Reynolds numbers greater than 100
\citep{ode87}, it is unlikely that the present phase of the IVC infall
will result
in triggered star formation. Triggered star formation will, however,
become more likely in the future as the
cloud moves from the low-density, high-temperature
environment of the halo into the high-density, low-temperature \ion{H}{I}
medium in the Galactic plane.
There is evidence for this transition in
IRAS 100 $\mu$m imaging of other cometary clouds found at
$|b| > 15^\circ$ \citep{ode88}. The clouds whose morphologies are
suggestive of supersonic flow and also supporting possible star formation are
preferentially at lower Galactic latitude and thus likely closer to the
Galactic plane.  Three of the cometary clouds studied by \citet{ode88},
G139-65, G192-67, and G225-66, are at high Galactic latitude.

\subsubsection{MBM 55:}

\begin{figure}[htpb]
\centering
\includegraphics[draft=False,width=\textwidth]{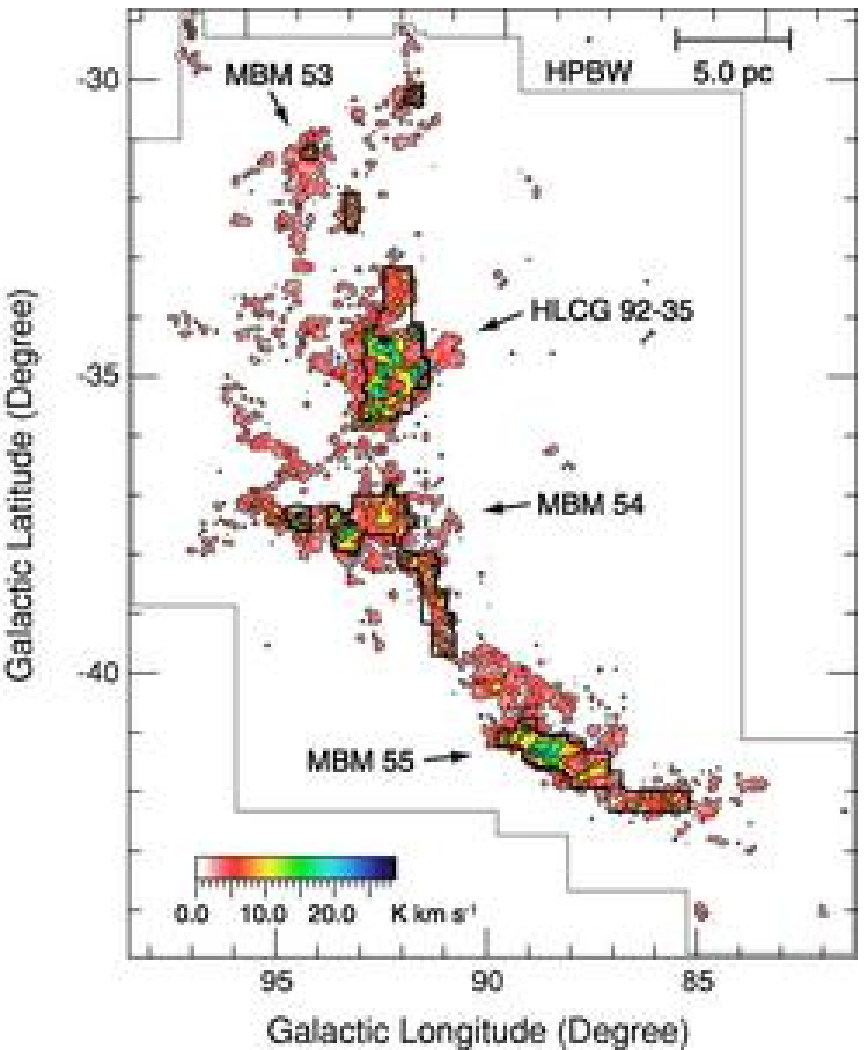}
\caption{Total velocity-integrated intensity  ${}^{12}$CO (J = 1-0) map of
MBM 53-55 in Galactic coordinates from \citet{yam03}. The lowest contour is
1.50 K km s$^{-1}$, and separation between the contours is 6.0 K km s$^{-1}$.
The observed area of ${}^{12}$CO is denoted by a thin solid line, and that of
${}^{13}$CO is denoted by thick solid lines.
1RXS J231019.1+144711, the WTTS candidate identified by \citet{li00}, is
located just East of the main core at $(l,b)=(89.51,-41.43)$.
\label{fig-mbm55}}
\end{figure}

MBM 55 $(l=89.19, b=-40.94)$ is at an estimated distance of 150 pc and is
the largest high latitude translucent
cloud, with a mass of 164 M$_\odot$ under the assumption that ${}^{13}$CO is in LTE
\citep{yam03}.
It contains high density cores that are evident in
CS(2-1) emission. This cloud is embedded along with MBM 53, HLCG 92-35, and
MBM 54 in a massive $\sim$1200 M$_\odot$ \ion{H}{I} arc spanning 15 degrees
on the sky (Figure \ref{fig-mbm55}). \citet{yam03} find that the CO/\ion{H}{I}
ratio varies by a factor of $\sim$10 across the arc indicating that the
hundreds of cloudlets that comprise the arc are in different molecular stages of
molecular cloud evolution. This structure is similar in morphology to the
Polaris Flare \citep{hei90}.

Spectroscopic followup of RASS selected objects
by \citet{hea99} and \citet{li00} yielded three possible PMS stars
projected near the core of MBM 55.
On the basis of the weak \ion{Li}{I} absorption exhibited by
1RXJ 2253.0+1650 and 1RXJ 2305.1+1633,
the two stars in their sample that were above the main sequence,
\citet{hea99} concluded that
both are luminosity class III or IV post-main sequence stars. As noted above,
\citet{hea99} interpret the majority of X-ray bright stars in this field and
towards MBM 7 as part of the local $\sim$150 Myr old population.
\citet{li00}, however, identified 1RXS J231019.1+14471 as a candidate WTTS
in spite of EW(Li) = $-$0.10{\AA},
based on the prediction of PMS models that early M stars tend to deplete
their lithium on the order of 10$^6$ years.

\subsection{Moving Groups and Associations}

Several of the pre-main-sequence stars found at $|b| > 30^{\circ}$
are members of local moving groups and stellar associations.
The chapter on Small Loose Associations by Torres et al.
in this Handbook discusses each of these groups in detail. It
is worthwhile noting that high-latitude PMS stars are found in the
AB Doradus moving group (HD 25457 aka HIP 18859), the kinematically similar
but older Local Subgroup B4 (BD+21 418),
the TW Hydrae association (HD 98800 aka TWA 4A), and
the $\beta$ Pictoris association (GJ 3305). These stars are within 55 pc,
and thus are seen at high Galactic latitude due to their proximity to
the Sun.

\subsection{Isolated PMS Objects}

\subsubsection{HD 141569:}

\citet{wei00} identify the triple system HD 141569 as a $\beta$ Pictoris-like
B9.5Ve primary with comoving M2V and M4V companions separated by
1.4$''$ and located at distances of 7.55$''$ and 8.93$''$, respectively.
The estimated age based on model
isochrones for the two low-mass WTTS is $5\pm3$ Myr, assuming they are at
the 100 pc distance of HD 141569 determined by Hipparcos.

This system is projected near the high latitude clouds MBM 34-39 and
the presence of interstellar absorption lines in the spectrum of
HD 141569A; \citet{sah98} conclude that it lies behind MBM 37. \citet{wei00}
note that HD 141569 may not be physically associated with these
clouds citing LSR radial velocity differences of $\sim$18 km s$^{-1}$
and the lack of infall signatures in the clouds. \citet{mar05} find
evidence in FUSE spectra for the diffuse outer region of L134N as well as a
turbulent region of the dark cloud (as deduced from excitation of
high-$J$ levels in H$_2$) being present along the sightline to HD 141569A.
The $UVW$ space motion of
the HD 141569 system is within two sigma of several nearby star formation
associations ($\eta$ Cha, TW Hya, and Tucana) suggesting
a related origin.

As one of the closest Herbig AeBe stars, HD 141569A and its debris disk
have been subjects of ongoing and extensive study.
In their Chandra ACIS-I survey of 17 Herbig AeBe stars \citet{ste06}
find that HD 141569A is one of 4 that do not exhibit X-ray emission,
with log($L_X/L_*$) $<$ $-$6.8. The low-mass companions
HD 141569B and C, however, are both
X-ray active having log($L_X/L_*$) = $-$3.3.
HD 141569 possesses a transitional disk which
\citet{hal06} failed to detect polarized emission from in $JHK$
using the IRPOL2+UIST instrument at UKIRT, although
the circumstellar disks around TW Hya, HD 169142, HD 150193, and
HD 142666 were clearly seen.
They attribute this to a relatively low peak surface
brightness of $0.2\pm1.2$ mJy arcsec$^{-2}$ for the HD 141569 disk
versus $5.7\pm1.4$ mJy arcsec$^{-2}$ for that of TW Hya.

The disk of HD 141569A exhibits a complex structure consisting of
an inner clearing and two spiral rings. On the basis of dynamical
simulations \citet{ard05} propose that the disk is in a transient
dynamical state due to a recent ($\sim$4000 yr) flyby of the two
M dwarf companions. The event is also invoked to explain the observed
dust creation rate exceeding that expected by submillimeter mass
of the disk \citep[0.33 M$_\odot$;][]{rhe07}.
\citet{got06} resolved the circumstellar disk using Subaru
AO+IRCS observations in CO $v$=2-1 finding that the inner clearing is
11$\pm$2 AU in radius, which they
interpret as close to the gravitational radius of the star, i.e. where
the sound speed in the ionized medium equals the escape velocity
of the system $r_g = GM_*/c^2_i$.

From an analysis of the Br$\gamma$ line profile \citet{gar06} infer a mass
accretion rate of $4.3\times10^{-9}$ M$_\odot$/yr. This accretion rate
is surprising given the large inner hole and the apparent lack of
circumstellar material \citep{gui06}. As HD 141569A is a rapid rotator
($v_* sini = 236$ km s$^{-1}$) \citet{bri07} propose that the emission
may originate from a compact circumstellar disk formed by ejected material.

\subsubsection{BP Psc:}

\citet{cor97} classify Herbig Ae/Be stars into four categories
based on the [\ion{O}{I}] $\lambda$6300 measured velocities and line profiles:
I - strongly blueshifted emission, accompanied by a lower velocity
blueshifted component, II - emission similar to the lower velocity
component in category I, III - low velocity redshifted emission, and
IV - unshifted and symmetric emission. BP Psc (aka StHA 202, PDS 103)
exhibits low velocity but broad blueshifted emission placing it
in category IIb, which are interpreted as Herbig Ae/Be
stars having evolved past the outflow stage.
BP Psc has the lowest line velocities of any star in
their sample that exhibits [\ion{S}{II}] emission.

The spectral type of BP Psc has been estimated as mid-F by
\citet{dow88} and more recently as G2IVe based on spectra obtained from the
HARPS echelle spectrograph at the ESO La Silla 3.6m telescope as part
of the variable star one-shot project (VSOP) \citep{dal07}.
\citet{sua08} also assign a spectral type of G2e to BP Psc as
part of their follow-up
observations of objects having IRAS FIR colors similar to
planetary nebulae.

However, due to an extensive multiwavelength observation program by
\citet{zuc08} the PMS status of BP Psc is now in doubt. In spite of
the presence of emission line
bi-polar jets and evidence for gas
accretion, \citet{zuc08}
suggest that BP Psc may be a first-ascent, post-main sequence
giant star whose accretion and outflows are due to a low-mass
companion possibly consumed
during a recent common envelope phase.

The list BP Psc's intriguing features include ${}^{12}$CO(2-1) and
${}^{12}$CO(3-2) emission, which are rare in isolated high Galactic
latitude main sequence and PMS stars and unknown in first-ascent giant
stars, an extremely compact circumstellar disk ($<$~0.2$"$), and a
very large fraction ($\sim75\%$) of the stellar bolometric luminosity
being reprocessed by grains. The optical and near-IR spectral features
include those consistent with a PMS nature such as a
broad (EW = -11 to -15 {\AA}) H$\alpha$ emission line,
numerous forbidden emission lines, and a $v$sini of 32.3 $\pm$ 1.2 km s$^{-1}$
suggestive of youth.
However, the EW of the lithium 6709.6 {\AA} line is only 50 m{\AA}, which
is more appropriate for K-type stars older than 100 Myr. In addition,
analysis of surface gravity sensitive features indicate log(g) $\sim$ 2.5,
an order of magnitude weaker than that expected for a PMS star.

\citet{zuc08} conclude that uncovering the true nature of BP Psc
will requre a direct measurement of its trigonometric parallax in order
to determine its luminosity. If BP Psc is a PMS star, then the
challenges of the apparent low surface gravity and lack of
photospheric lithium must be dealt with.
In the alternative scenario BP Psc may be the first example of a rare stage
in the post-main sequence evolution of close binary systems.

\subsubsection{GSC 04744-01367:}

Both members of this 6$''$ wide binary, cataloged as PDS 11a/b by
 \citet{gre92}, exhibit H$\alpha$ emission, although Li absorption
is only seen in the northern (a) component. No spectral type is
specified although the photometry given by \citet{gre92}
($B-V = 1.51, 1.34$ and $V-I = 2.21, 2.24$) for the two components are
consistent with an
early M spectral type \citep{rei00}. The minimum photometric distances
computed using the main sequence $(M_V,V-I)$ relation in \citet{rei00} and
based on the $V-I$ colors and $V$ magnitudes of $V = 14.76, 15.34$ are 77 and
97 pc, respectively.

\subsubsection{TYC 8474-24-1:}
TYC 8474-24-1, aka PDS 2 \citep{gre92}, is an
isolated Herbig Ae/Be star of spectral type F3V \citep{vie03}.
The photometric distance as a main sequence star is 340 pc,
so at
$b = -64.1^{\circ}$ the star's minimum height below the Galactic plane
is 306 pc.

\section{Discussion: Environment and Processes of High-latitude Star Formation}

There are three distinct populations of pre-main-sequence stars
seen at high Galactic latitudes. These are those still found in association
with their parent molecular clouds, those found in isolation far from
any present molecular cloud, and those formed in extraplanar environments
high above the Galactic plane.

The processes that result in isolated T Tauri stars include ejection of the
young star and dissipation of the parent cloud. If dynamical ejection
velocities are as high as $\sim$5 km/sec, over the typical 10 Myr
lifetime of the Classical T Tauri phase a star may have traveled 50 pc
resulting in separations from their birthplace by 10's of degrees on the
sky (for distances of a few 100 pc).

The known high-latitude star formation sites include both gravitationally
bound (MBM 20) and unbound (MBM 12) clouds. When star formation occurs
within an unbound cloud the cloud dissipates on timescales of a few Myr,
even without massive O/B stars present. It is possible that outflows
from low-mass stars may drive the dissipation \citep{ste95, fei96, har01}.

\subsection{Determining Association between PMS Stars and Clouds}

The example of HD 141569 illustrates the difficulty of associating PMS
candidates with high-latitude clouds. There are three major issues
determining the association between a PMS star and a cloud.
The first is that candidate PMS stars are seldom projected
against the cores of the clouds. This is in contrast to IRAS studies
of PMS stars associated with dark cloud cores, e.g. \citet{woo94},
and
forces us to make proximity arguments.
The second issue is that the spatial velocity of the candidate PMS star
can be very discrepant with that of the gas. Models of T Tauri migration
predict velocity dispersions of  $\sim$1 km s$^{-1}$ inherited from
thermal motions in the parent cloud \citep{fei96}. While dynamic ejection
can, as stated above, result in velocity dispersions an order of magnitude
greater, this process may be less effective in the lower density
environments of the high latitude star formation regions.

The third complication is that the distances to both the cloud and to the PMS
star are usually poorly determined. Cloud distances can be inferred by
techniques such as comparison of
off-cloud and on-cloud star counts, bracketing of cloud location by the
detection of reddening in stellar colors and absorption features in the
spectra of early type stars, and
by statistical estimates based on an assumed dust scale height above
the Galactic plane.

The luminosity of a PMS star as it descends down its
nearly isothermal Hayashi track has a steep dependency on the
stellar age and mass \citep{bar98}. In order to accurately determine the
intrinsic luminosity of a PMS star, and thus its distance, the photometry
and spectra must be analyzed and compared against evolutionary models.

\subsection{High-latitude Stellar Nurseries}

The low density environment found in high-latitude clouds
creates a challenge for low-mass star formation via direct
gravitational collapse.
One scenario is that stars are created at the low density limit of
the normal formation process, i.e. gravitational collapse of
low mass cores.
There are a handful of star formation regions found in
high-latitude molecular clouds,
e.g. in MBM 12 and MBM 20.

Both MBM 12 and MBM 20 are Lynds dark clouds thus
the column density of H$_2$ in their densest regions is higher by
an order of magnitude than for typical translucent high-latitude clouds.
We suspect these are the closest
star formation regions to the Sun - besides \ion{Na}{D} studies,
color excess, etc., in a statistical sense we know these
predominately translucent clouds are nearby because they
are more numerous in the southern Galactic hemisphere (the
Sun is $\sim$14 pc north of the midplane).

Until recently MBM 12 was considered the nearest star-forming region,
however recent work \citep{luh01,and02} has placed it at $\sim$275 pc,
so that it is significantly
farther out than the Taurus or Ophiuchus star forming regions.
MBM 20 may now be the closest star forming region, but its distance is not
well-constrained, with estimates ranging from 100 to 160 pc.

\subsection{Isolated High-latitude PMS Stars}

High-latitude PMS stars found in isolation may be the result of
either ejection from or dissipation of the parent cloud.
Dissipation timescales for unbound clouds appear to be a few
Myr.
The formation scenarios for low-mass isolated T Tauri stars include normal
modes of formation within a molecular cloud followed by dissipation
of the parent cloud and/or dynamical ejection of the young
star.
The population of dispersed Weak-lined T Tauri stars (WTTS) is
doubtless underestimated since due to the absence of magnetospheric
accretion and disk signatures they are nearly indistinguishable
from the numerous active M dwarfs (dMe) in the field.

\subsection{Extraplanar Star Formation Regions}

\begin{figure}[tb]
\includegraphics[draft=False,width=\textwidth]{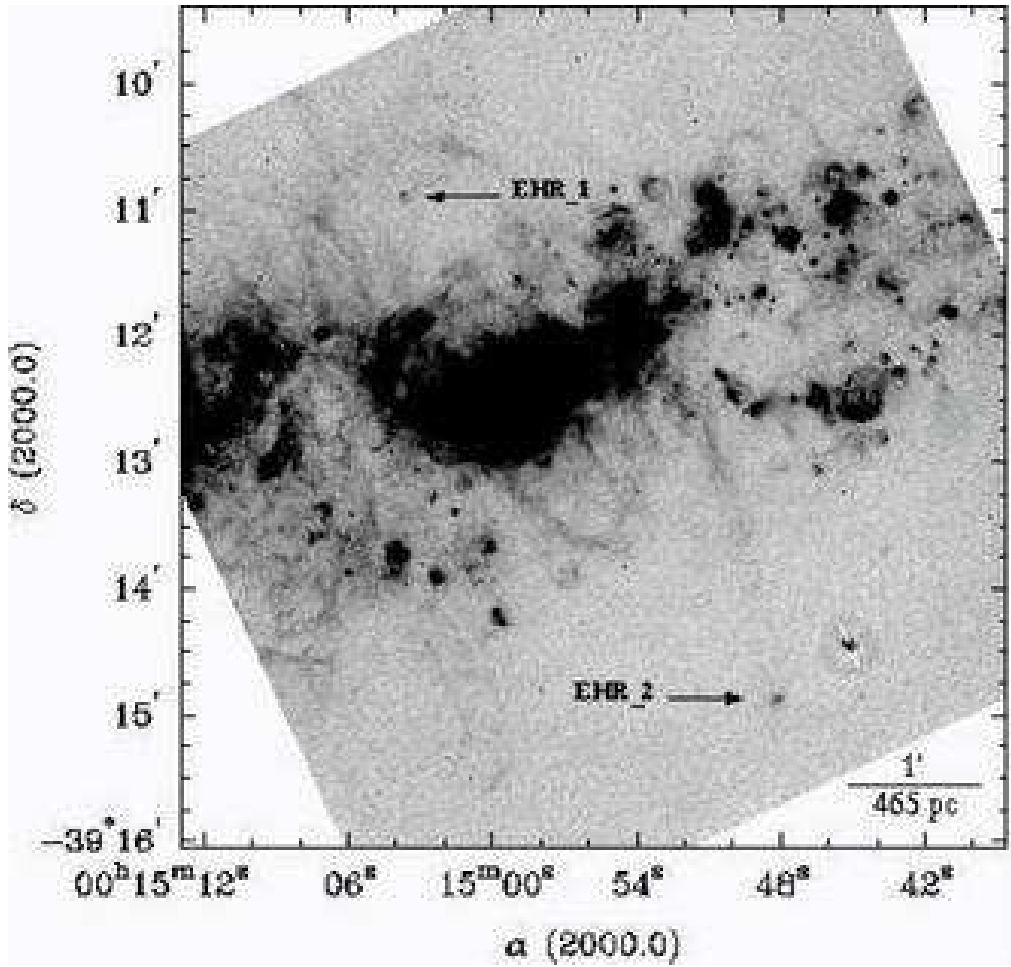}
\caption{This H$\alpha$ image from the VLR FORS instrument of the nearby
edge-on spiral galaxy NGC 55 \citep{tue03} shows two extraplanar
H II regions. Hydrodynamic considerations suggest that the high-mass
star formation occurred in-situ.
\label{fig-ngc55}}
\end{figure}

In the Second and Third Galactic quadrants we are afforded a view of
the disk-halo interface relatively free of interstellar extinction and
obscuration by the Galactic bulge population. This provides the
opportunity of an optical survey of stellar populations formed in-situ
in the Galactic halo. In nearby edge-on spirals, for example NGC 55
in the Sculptor group, there is clear evidence for OB associations found
high above the Galactic disk \citep[][see Figure \ref{fig-ngc55}]{tue03}.
Hydrodynamic considerations
suggest that the high-mass star formation occurred in-situ.
The signatures of Galactic high-latitude massive star formation include
\ion{H}{II} regions \citep{sha59,gil77,bli82} that would be detected via
imaging and H$\alpha$ surveys, e.g. \citet{rey05}.

The interpretation of these objects, however, can be problematic as
shown by examination of the presumed \ion{H}{II} regions cataloged by
\citet{sha59}.  Out of the 8 high-latitude Sharpless regions, 2 are
planetary nebulae and the remainder are dust reflection nebulae or
photoluminescence of C in molecular clouds, thus none are bona fide
\ion{H}{II} regions.  The specific identifications for the planetary
nebulae are PN A55 20 and PN A55 24 \citep{abe55} for Sh2-290 and
Sh2-313.  Sh2-24, Sh2-33, Sh2-36, Sh2-73, Sh2-122, and Sh2-245, which
are the other Sharpless regions at $|b| \ge 30^{\circ}$, are associated
with the molecular clouds MBM 57, MBM 38, MBM 38, MBM 40, MBM 55, and
MBM 18, respectively \citep{bli82,mag85}.

\citet{nym87} find a large molecular cloud in Lupus that is far from the
Galactic plane. At an assumed distance of 2.9 kpc, this cloud is 150 to 280 pc
above the plane. The relative lack of CO emission in the plane
below the cloud, with a diameter of $\sim$200 pc, suggest that
stellar winds and/or supernovae may have formed both structures.
\citet{nym87} point out a semblance to the local Gould Belt and a relation,
albeit on different scales, to \ion{H}{I} shells and Galactic fountains.

A possible formation scenario involves a star formation episode within
the disk ejecting gas into the halo. This gas subsequently cooled down
forming \ion{H}{I} and possibly molecular clouds. Winds and shocks from
later disk star formation episodes then trigger star formation
in these clouds \citep{elm98}. Presumably low-mass star formation would
be triggered as well although these would not be detected in H$\alpha$
surveys of external galaxies such as \citet{tue03}.
This association of triggered star formation within the halo
with shocked gas from superbubbles and Galactic fountain events
is relevant to the search for stellar populations in Intermediate and
High Velocity Clouds (IVCs and HVCs).
Moreover, if stellar populations are indeed formed within the halo from
disk gas transported out of the plane by massive star winds and supernovae,
they should combine near-solar metallicities with non-disk kinematics due to
ejection and infall.

Deep and high-resolution optical imaging of massive (L$_*$) edge-on spiral
galaxies reveals that extraplanar $(|z| >$ 0.4 kpc) dust structures are
found in association with extraplanar H$\alpha$ emission
\citep{how99, tho04}. Both the dust and
the H$\alpha$ emission are seen above most of the thin disk molecular
material. The largest structures have gas masses $> 10^5$ M$_\odot$ with
implied potential energies $> 10^{52}$ erg.
The morphologies of some of the extraplanar dust structures
suggest production via supernova-driven galactic fountain or chimney phenomena.
Other extraplanar features, for example as seen in NGC 4217,
are not readily linked to high-energy events in the disk.

\section{Summary}

The high latitude population of molecular clouds is mostly translucent.
Although molecular cores have been identified in translucent clouds and even
in Galactic cirrus, the only verified stellar nurseries are
MBM 12 and MBM 20, which are both dark clouds.
Thus, at the very least, the star formation rate in translucent clouds is
significantly less than in dark clouds.
Is this a consequence of column density, which is the
most obvious distinction between dark and translucent clouds, rather than
density within the cloud?

The ability of at least a few high latitude clouds to form cold molecular
cores and young stars may be due to a combination of conditions including
variations in the interstellar radiation field, changes in dust grain
size and chemistry, and the occurrence of shocks and other transient
events in the ISM. Regardless of the formation mechanisms, study of the
high latitude star formation environment continues to be an intriguing
field.

\acknowledgements
I am thankful to  Loris Magnani for a detailed and helpful referee's
report.
This research has made extensive use of the SIMBAD database, operated at CDS,
Strasbourg, France, and NASA's Astrophysics Data System.

\end{document}